\documentclass[sigconf]{acmart}
\acmConference[ICSE 2022]{The 44th International Conference on Software Engineering}{May 21–29, 2022}{Pittsburgh, PA, USA}

\copyrightyear{2022} 
\acmYear{2022} 
\setcopyright{acmcopyright}\acmConference[ICSE '22]{44th International Conference on Software Engineering}{May 21--29, 2022}{Pittsburgh, PA, USA}
\acmBooktitle{44th International Conference on Software Engineering (ICSE '22), May 21--29, 2022, Pittsburgh, PA, USA}
\acmPrice{15.00}
\acmDOI{10.1145/3510003.3510133}
\acmISBN{978-1-4503-9221-1/22/05}

\usepackage[utf8]{inputenc}

\usepackage{natbib}
\usepackage{graphicx}
\usepackage{xspace}
\usepackage{pifont}
\usepackage{caption}
\usepackage{subcaption}
\usepackage{multicol}
\usepackage{enumitem}
\usepackage{multirow}
\usepackage{tcolorbox}
\usepackage{xcolor}

\usepackage{listings}
\usepackage[linesnumbered,ruled,noend,vlined]{algorithm2e}
\usepackage{setspace}

\usepackage{hyperref}
\hypersetup{
    colorlinks=true,
    linkcolor=blue,
    filecolor=magenta,      
    urlcolor=cyan,
}

\usepackage{listing}
\newcommand*{\tool}{{\textsc{Morest}}\xspace}
\newcommand*{\toolatmonly}{{\textsc{Morest-RPG-ONLY}}\xspace}
\newcommand*{\toolnoatm}{{\textsc{Morest-NO-RPG}}\xspace}

\newcommand*{\atm}{{{RESTful-service  Property Graph}}\xspace}
\newcommand*{\atmab}{{{RPG}}\xspace}

\newcommand*{\rstapi}{{RESTful API}\xspace}
\newcommand*{\rstservice}{{RESTful service}\xspace}
\newcommand*{\restler}{{\textsc{Restler}}\xspace}
\newcommand*{\restgen}{{\textsc{Resttestgen}}\xspace}
\newcommand*{\evomaster}{{\textsc{Evomaster}}\xspace}
\newcommand*{\evomasterbb}{{\textsc{Evomaster-bb}}\xspace}

\newtheorem{myDef}{Definition}

\newcommand{\yi}[1]{\textcolor{red}{Yi: #1}}

\newcommand{\header}[1]{{\smallskip \noindent
\textbf{#1}}\xspace}

\newcommand{\listitem}[1]{\noindent\textbf{#1}\xspace}

\newcommand*{\figu}{{Fig.}\xspace}
\newcommand*{\defi}{{Def.}\xspace}
\newcommand*{\sect}{{Section}\xspace}
\newcommand*{\tabl}{{Table}\xspace}

\newcommand*{\algo}{{Algo.}\xspace}

 \usepackage{microtype}
    \makeatletter
\def\@fnsymbol#1{\ensuremath{\ifcase#1\or \dagger\or \ddagger\or
   \mathsection\or \mathparagraph\or \|\or **\or \dagger\dagger
   \or \ddagger\ddagger \else\@ctrerr\fi}}
    \makeatother

\begin{document}

\title{\tool{}: Model-based RESTful API Testing with Execution Feedback}

\author{Yi Liu}
\affiliation{
  \institution{Nanyang Technological University}
  \country{Singapore}}
  
\author{Yuekang Li}
\authornotemark[1]
\thanks{{$\dagger$}Corresponding author}
\affiliation{
  \institution{Nanyang Technological University}
  \country{Singapore}}
  
\author{Gelei Deng}
\affiliation{
  \institution{Nanyang Technological University}
  \country{Singapore}}
  
\author{Yang Liu}
\affiliation{
  \institution{Nanyang Technological University}
  \country{Singapore}}

\author{Ruiyuan Wan}
\affiliation{
  \institution{Huawei Cloud Computing Technologies Co., Ltd}
  \country{China}}

\author{Runchao Wu}
\affiliation{
  \institution{Huawei Cloud Computing Technologies Co., Ltd}
  \country{China}}

\author{Dandan Ji}
\affiliation{
  \institution{Huawei Technologies Co., Ltd}
  \country{China}}

\author{Shiheng Xu}
\affiliation{
  \institution{Huawei Cloud Computing Technologies Co., Ltd}
  \country{China}}

\author{Minli Bao}
\affiliation{
  \institution{Huawei Cloud Computing Technologies Co., Ltd}
  \country{China}}

\begin{abstract}

RESTful APIs are arguably the most popular endpoints for accessing Web services.
Blackbox testing is one of the emerging techniques for ensuring the reliability of RESTful APIs.
The major challenge in testing RESTful APIs is the need for correct sequences of API operation calls for in-depth testing.
To build meaningful operation call sequences, researchers have proposed techniques to learn and utilize the API dependencies based on OpenAPI specifications.
However, these techniques either lack the overall awareness of how all the APIs are connected or the flexibility of adaptively fixing the learned knowledge.

In this paper, we propose \textsc{Morest}, a model-based RESTful API testing technique that builds and maintains a dynamically updating RESTful-service Property Graph  (RPG) to model the behaviors of RESTful-services and guide the call sequence generation.
We empirically evaluated \textsc{Morest} and the results demonstrate that \textsc{Morest} can successfully request an average of 152.66\%-232.45\% more API operations, cover 26.16\%-103.24\% more lines of code, and detect 40.64\%-215.94\% more bugs than state-of-the-art techniques.
In total, we applied \textsc{Morest} to 6 real-world projects and found 44 bugs (13 of them cannot be detected by existing approaches). Specifically, 2 of the confirmed bugs are from Bitbucket, a famous code management service with more than 6 million users.

\end{abstract}

\maketitle

\section{Introduction}


Representational state transfer (REST) has become a de-facto standard for Web service interactions since it was introduced in 2000~\cite{rest-root}.
Web-based APIs which follow this standard are called \rstapi{}s and the web services providing the \rstapi{}s are called \rstservice{}s.
Nowadays, most web service providers, such as Google~\cite{google}, Twitter~\cite{twitter} and Amazon~\cite{amazon}, expose \rstapi{}s to grant access to other applications or services.
As the \rstapi{}s gain popularity, techniques for automatically testing them become important.
Based on whether the knowledge of the program internals is needed or not, these testing techniques can be categorized as whitebox and blackbox techniques.
Whitebox techniques are normally more effective but require source code~\cite{evomaster}.
On the contrary, blackbox techniques only rely on a well-defined interface to conduct testing~\cite{restler, resttestgen}.
In comparison to their whitebox counterparts, blackbox techniques enjoy superior applicability considering that a cloud service can be implemented with different programming languages and may use third-party libraries whose source code is not available.
In this paper, we concentrate on the blackbox \rstapi{} testing techniques.



One of the most challenging problems in testing \rstapi{}s is how to infer the correct sequences of calling the API operations (aka, \emph{call sequences}) where each API operation can be one of the four basic types --- create, read, update, and delete (CRUD).
This is because \rstapi{}s are often organized sparsely to encapsulate different micro services and fulfilling a single task can involve a chain of API calls.
Take the Petstore service~\cite{swagger-petstore} as an example, it is a \rstservice{} for selling and ordering pets.
In Petstore, before calling the API to order a pet, APIs for creating the pet and updating its status as ``available'' must be invoked.
Skipping any of the prerequisite APIs will cause the ordering pet operation to fail, preventing the coverage of deeper logic in the code.
To address the challenge of generating proper API call sequences, researchers have proposed several testing techniques~\cite{ed2018automatic, restler, resttestgen} which can infer the dependencies between \rstapi{}s to guide the test generation.
For the purpose of API dependency inference, these techniques leverage API specifications such as OpenAPI~\cite{openapi-spec}, RAML~\cite{raml-spec} and API Blueprint~\cite{apiblueprint-spec}.
Among these API specifications, OpenAPI (aka, Swagger) is becoming increasingly popular and gets adopted by major IT companies (e.g., Google, Microsoft, and IBM).

Two most recent state-of-the-art blackbox \rstapi{} testing techniques --- \restler{}~\cite{restler} and \restgen{}~\cite{resttestgen} use the OpenAPI specifications of the target \rstservice{}s to facilitate their call sequence generation.
On the one hand, both of them learn the \textit{producer-consumer} dependencies~\footnote{If a resource in the response of an API A is used as an input argument of another API B, then B depends on A.} between the APIs to enforce the correct ordering of APIs in the generated call sequences.
On the other hand, the difference between \restler{} and \restgen{} lies in how they utilize the learnt dependencies:
\ding{182} For \restler{}, it uses a \textit{bottom-up} approach, which starts with testing single APIs and then extends the API call sequences by heuristically appending API calls.
Although \restler{} can limit the search space with dynamic feedbacks (i.e., if certain combination of APIs fails to execute, \restler{} avoids this pattern in the future), the search space for extending the test sequences is still very large due to the lack of overall awareness of how the APIs are connected.
\ding{183} To gain such awareness, \restgen{} proposes a \textit{top-down} approach to connect the APIs into an \textit{Operation Dependency Graph} (ODG), where APIs are nodes and their dependencies are edges.
With the ODG built, \restgen{} can then traverse it and aggregate the visited API nodes to generate call sequences.
In this sense, \restgen{} can generate valid call sequences more efficiently.
However, the quality of the tests generated by \restgen{} might be hindered since it heavily depends on the ODG, which may not reflect the API behavior correctly due to some pitfalls (e.g., poorly written OpenAPI specifications).
In short, both the \textit{bottom-up} and \textit{top-down} approaches have strengths and weaknesses, leaving the generation of proper API call sequences an under-researched field.

In this paper, we propose \tool{}
-- a blackbox \rstapi{} testing technique with a dynamically updating \atm{} (\atmab{}).
The workflow of \tool{} contains two major procedures: building the \atmab{} with the OpenAPI specifications of the target \rstservice{} and the model-based testing with dynamic updates of the \atmab{}.
The \atmab{} is a novel representation of \rstservice{}s proposed in this paper which encodes 
API and object schema information in the form of a mixed, edge-labeled, attributed multigraph.
Compared to ODG, \atmab{} can model not only the producer-consumer dependencies between APIs with more details but also the property equivalence relations between schemas, which allows \atmab{} to both describe more behaviors of the \rstservice{}s and flexibly update itself with execution feedback.
By traversing the \atmab{}, \tool{} can aggregate the visited APIs to build meaningful call sequences.
During the testing process, \tool{} constantly collect the responses of the \rstservice{} and use the dynamic information to update the \atmab{} adaptively.
With the updated \atmab{}, \tool{} can then generate test sequences with better quality for the next iteration of testing.
In this sense, \tool{} enjoys the benefits of both the \textit{bottom-up} and \textit{top-down} approaches while avoiding their drawbacks by having both the overall awareness of all APIs and the flexibility of making changes.

We empirically evaluated \tool{} on six \rstservice{}s running in local environment. 
In our experiments, \tool{} outperforms the state-of-the-art blackbox \rstapi{} testing tools, namely \restler{} and \restgen{} with superior average code coverage (26.16\% and 103.24\% respectively), average successfully requested operations (152.66\% and 232.45\% respectively) and average number of detected bugs (40.64\% and 111.65\% respectively).
In total, we applied \textsc{Morest} to six real-world projects and found 44 bugs (13 of them cannot be detected by existing approaches). In specific, 2 of the confirmed bugs are from Bitbucket, a famous Git code management service with more than six million users~\cite{bitbucket-user}.

\textbf{Contribution.} We summarize our contributions as follows:
\vspace{-1pt}
\begin{itemize}
    \item We propose a novel model called \atm{} (RPG) for describing Web services and adopt it for \rstapi{} testing.
    \item We develop a methodology for adaptively updating the RPG for enhanced performance.
    \item We evaluate the performance of \tool{} and demonstrate the superiority of \tool{} comparing to the state-of-the-art techniques with 1,440 CPU Hours.
    To the best of our knowledge, this is the first work to empirically compare blackbox \rstapi{} testing techniques.
    \item We detect 44 bugs with \tool{} in 6 projects.
    We responsibly disclose the bugs to the developers and 2 of them are confirmed until the time of writing this paper.
    \item We release our datasets and implementation of \tool{} to facilitate future research.
\end{itemize}

Currently, we have released the raw experimental data, the prototype of \tool{} and the evaluated benchmarks on the companion website of this paper.
The link to the website is \url{https://sites.google.com/view/restful-morest/home}.

\section{Background \& Running Example}

\subsection{Background} \label{sec:bg}

\header{\rstapi{}.}
The REpresentational State Transfer (REST) architecture is is first proposed by Roy Fielding in 2000~\cite{rest-root}.
A Web API using the REST architecture is called a \rstapi{}.
A Web service providing \rstapi{}s is called a \rstservice{}.
One of the fundamental constraints of REST architectural style is \textit{Uniform Interface}, which regulates the CRUD operations on the resources.
In modern Web API design practice, \rstapi{}s often use HTTP protocol as the transportation layer.
Therefore, the CRUD operations of \rstapi{}s can be mapped to the HTTP methods POST, GET, PUT and DELETE respectively.

\begin{figure}[t]
     \centering
     \begin{subfigure}[b]{0.4745\linewidth}
         \centering
         \includegraphics[width=\textwidth]{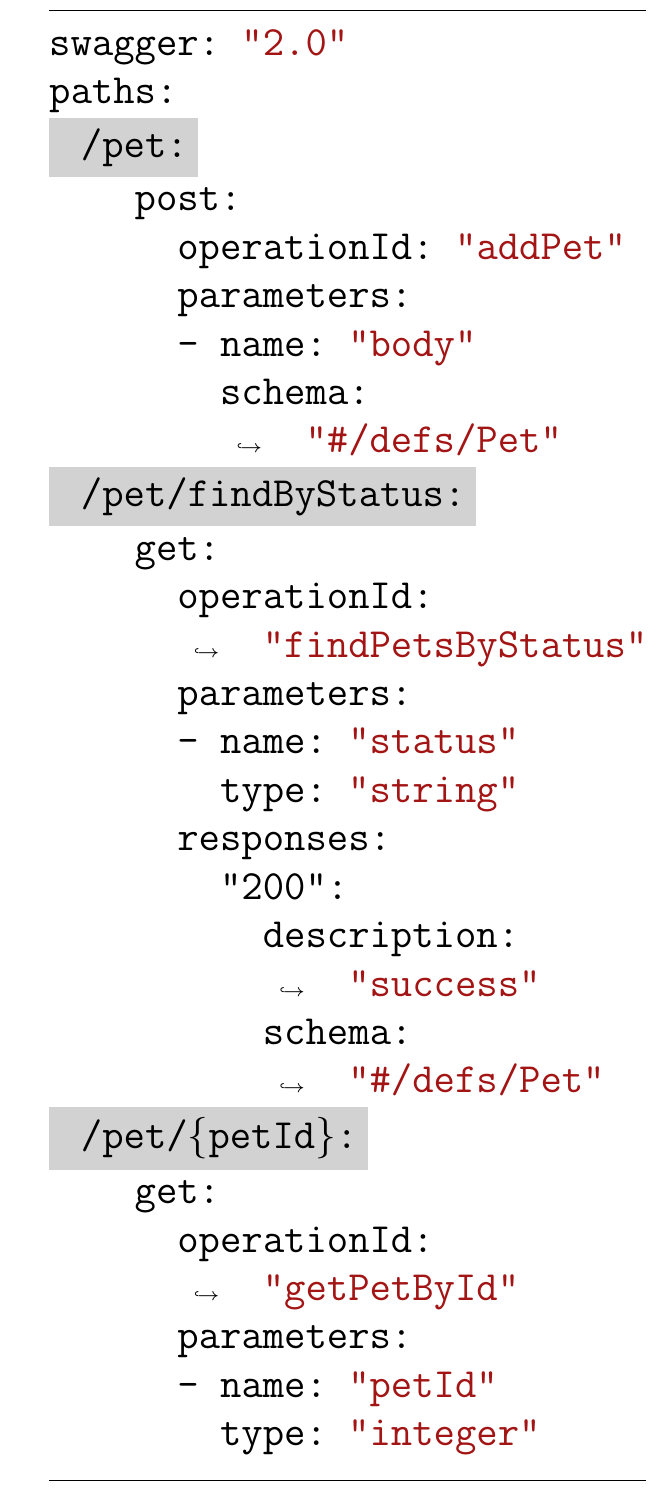}
         \caption{}
         \label{fig:api-def-a}
     \end{subfigure}
     \hfill
     \begin{subfigure}[b]{0.48\linewidth}
         \centering
         \includegraphics[width=\textwidth]{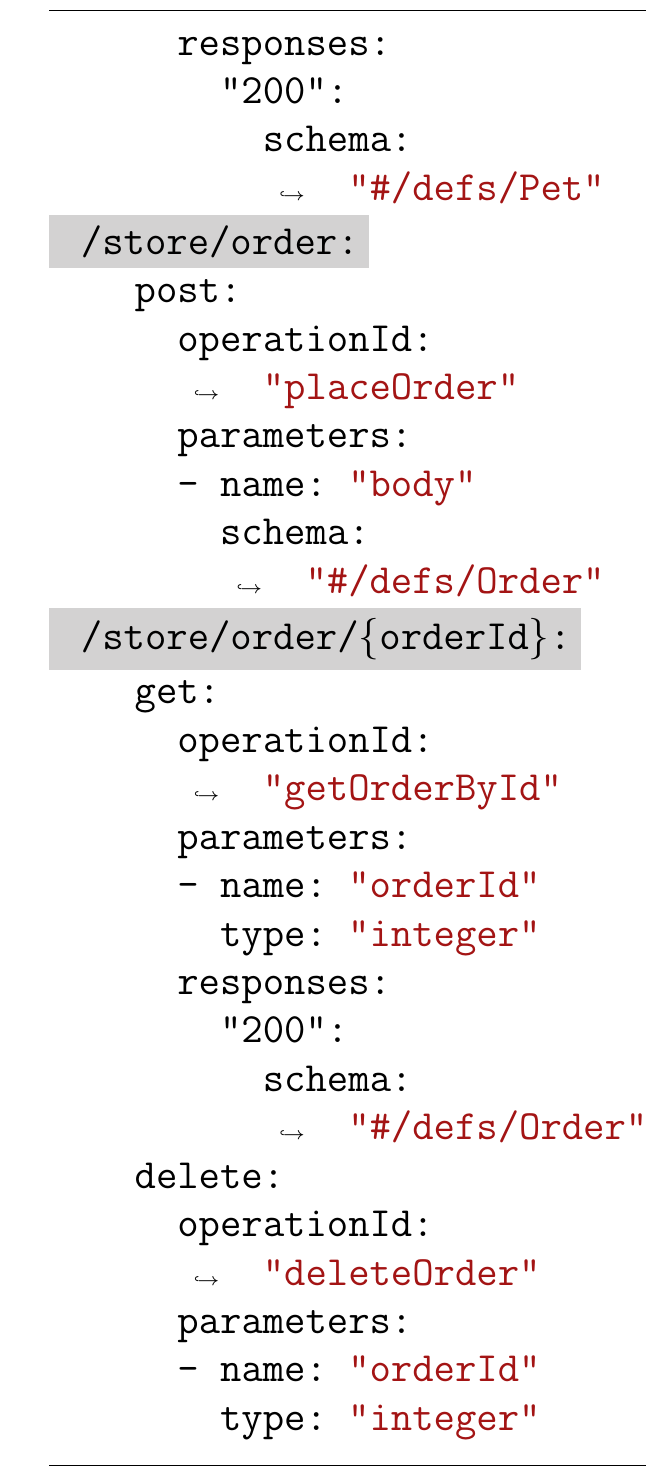}
         \caption{}
         \label{fig:api-def-b}
     \end{subfigure}
     \caption[Caption for Petstore \rstapi definitions]{The OpenAPI specification of Petstore APIs\textsuperscript{$*$}}
	\scriptsize{\textsuperscript{$*$} For clarity, we omit some details in the YAML file.}
	\label{fig:api-definition}
\end{figure}

\begin{figure}[t]
     \centering
     \begin{subfigure}[b]{0.4745\linewidth}
         \centering
         \includegraphics[width=\textwidth]{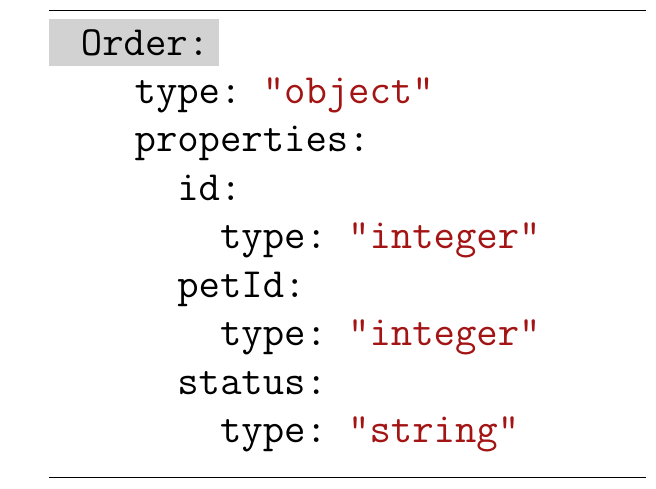}
         \caption{}
         \label{fig:schema-a}
     \end{subfigure}
     \hfill
     \begin{subfigure}[b]{0.48\linewidth}
         \centering
         \includegraphics[width=\textwidth]{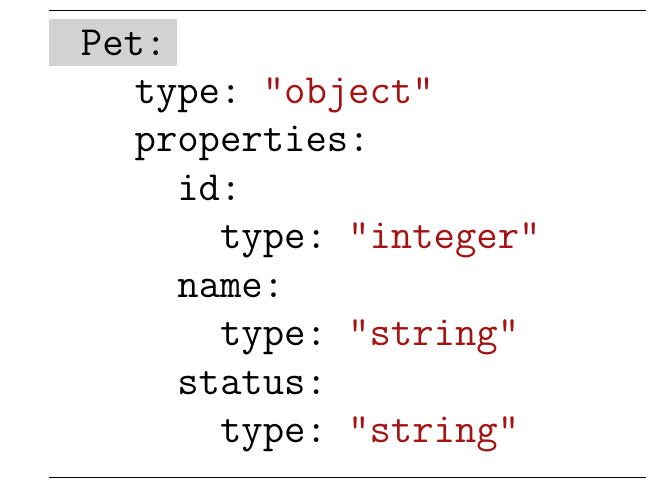}
         \caption{}
         \label{fig:schema-b}
     \end{subfigure}
     \caption[Caption for LOF]{The OpenAPI specification of Petstore Schemas \textsuperscript{$*$}}
	\scriptsize{\textsuperscript{$*$} For clarity, we omit some details in the YAML file.}
	\label{fig:schema-definition}
\end{figure}

\header{OpenAPI Specification.}
OpenAPI (previously known as Swagger) defines a standard for describing \rstapi{}s~\cite{openapi-spec} and an API document following this standard is called an OpenAPI specification.
OpenAPI specifications contain the information of the object \emph{schemas} as well as the API \emph{endpoints}, including but not limited to the available CRUD \emph{operations}, input parameters as well as responses.
These specifications can be stored as either YAML or JSON files.

\figu{}~\ref{fig:api-definition} shows a fragment of the OpenAPI specification for APIs in the Petstore service~\cite{swagger-petstore}.
In this example, five API endpoints are specified and they are marked with grey background.
We can see that each API endpoint supports one or more CRUD operations, which are specified by the \texttt{operationId} property.
In total, six operations are described in \figu{}~\ref{fig:schema-definition}, showing their input parameters and responses.
For an input parameter, it can be inside the request body (\texttt{body} of \texttt{addPet}) or in the URL path~\footnote{Some GET operations may involve the usage of query parameters and they are also considered as serialized in the URL path.} (\texttt{petId} of \texttt{getPetById}).
For a response, it contains the HTTP status code as well as the content body.
In addition, some operation parameters and responses may involve objects that are described by schemas.
For example, the operation \texttt{getPetById} returns responses with objects under the \texttt{Pet} schema.
\figu{}~\ref{fig:schema-definition} shows the OpenAPI specifications of the schemas.


\begin{figure*}[t]
	\centering
	\includegraphics[width=\linewidth]{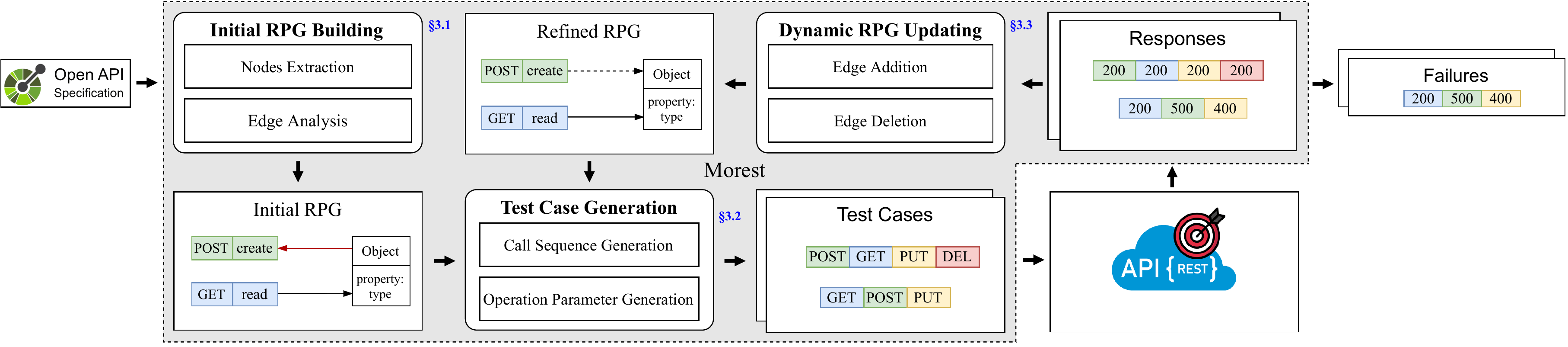}
	\caption{The workflow of \tool{}}
	\label{fig:work-flow}
\end{figure*}

\header{Operation Dependency Graph}
In \restgen{}, Viglianisi et al. propose to use Operation Dependency Graph (ODG) to model data dependencies among operations which can be inferred from the OpenAPI specifications~\cite{resttestgen}.
The benefit of using ODG is that aggregating operations by traversing the ODG can generate meaningful call sequences.
The ODG is a graph $G=(V, E)$ where $V$ is a set of nodes and $E$ is a set of directed edges.
For each node $v \in V$, $v$ corresponds to a unique operation in the OpenAPI specification.
The graph is said to contain a directed edge $e$ where $e = (v_2, v_1)$ for $v_1, v_2 \in V$ and $v_1 \neq v_2$, if and only if (iff) there exists a \textit{common field} in the response of $v_1$ and in the request parameter of $v_2$.
Two fields are \textit{common} when they have the same name if they are of atomic type (e.g., string or numeric) or when they are related to the same schema if they are of a non-atomic type.
Thus, following this definition, $v_2 \to v_1$ means $v_2$ depends on $v_1$.

\subsection{Running Example}

Here we introduce a running example extracted from the Petstore service~\cite{swagger-petstore}, which can demonstrate the limitations of existing approaches and aid in the explanation of \tool{}'s strategies in \sect{}~\ref{sec:method}.
\figu{}~\ref{fig:api-definition} and~\ref{fig:schema-definition} illustrate the OpenAPI specifications of the Petstore service.
The question is, given these specifications, how to generate meaningful API call sequences.
For this purpose, we need to infer the dependencies among the APIs.
Intuitively, the definition of API dependencies in ODG (\S~\ref{sec:bg}) can be applied here (both \restler{}~\cite{restler} and \restgen{}~\cite{resttestgen} follow this definition).
For the example in \figu{}~\ref{fig:api-definition}, we can get the following dependencies:
{\footnotesize
\begin{multicols}{2}
    \begin{enumerate}[leftmargin=*]
        \item \texttt{findPetsByStatus}$\to$\texttt{getPetById}
        \item \texttt{findPetsByStatus}$\to$\texttt{getOrderById}
        \item \texttt{addPet}$\to$\texttt{findPetsByStatus}
        \item \texttt{addPet}$\to$\texttt{getPetById}
        \item \texttt{getPetById}$\to$\texttt{getOrderById}
        \item \texttt{placeOrder}$\to$\texttt{getOrderById}
    \end{enumerate}
\end{multicols}
}
Note that some of the dependencies are infeasible: One example is in dependency 2 where \texttt{findPetsByStatus} should not depend on the \texttt{status} of the Order object returned by \texttt{getOrderById} because the status of a pet and the status of an order are not the same.
Another example is in dependency 4 where \texttt{addPet} should not depend on \texttt{getPetById} although both of them use the Pet schema.
While there are infeasible dependencies, they cannot be safely filtered out as yet because the current ODG model lacks the ability to describe the relation between an API and a schema in detail and dynamic execution feedbacks are needed to infer correct dependencies.

After the dependencies are acquired, we can then use them to guide the call sequence generation.
To start with, we can use a \textit{bottom-up} approach by testing single APIs first and then extend the test sequences by appending more APIs one at a time.
For example, we can start with a new sequence: \texttt{<getOrderById>}.
After a successful call of \texttt{getOrderById}, we can feed the \texttt{petId} of the returned Order object to \texttt{getPetById} according to dependency 5.
Then the sequence becomes \texttt{<getOrderById, getPetById>}.
Similarly, we can append \texttt{findPetsByStatus}  to the sequence according to dependency 1 and so on.
In this process, we can identify the infeasible dependencies and avoid using them in the future.
For instance, we can tell dependency 4 is faulty because if we try to append \texttt{addPet} to  \texttt{<getOrderById, getPetById>} according to dependency 4, the \texttt{addPet} operation will \emph{always} fail since the service refuses to create an already existing pet.
The limitation of the \textit{bottom-up} approach is that it lacks the overall awareness of how the APIs are connected with each other.
For example, it is hard to conduct Depth-First-Search (DFS) when extending a sequence because the potential length of the sequence is unknown and if it is very long, the risk of getting stuck in a local optimal is high.

Alternatively, we can use ODG to facilitate the generation of call sequences, which is a \textit{top-down} approach.
The call sequences can be generated by traversing the ODG and chaining the visited operations.
Compared with the \textit{bottom-up} approach, with ODG, we can quickly generate call sequences without trial and error.
However, because some extracted dependencies (such as 2 and 4) might be infeasible, the quality of the generated call sequences can be affected.
In other words, the \textit{top-down} approach lacks the flexibility of dynamically fixing the call sequences.

In summary, both the \textit{bottom-up} and \textit{top-down} approach have their own strengths and weaknesses and they can complement each other.
Therefore, a new model is needed to both provide high level guidance and perform self-updates.
Following this observation, we propose \tool{}, which leverages \atm{} (\atmab{}) to adopt the advantages of both approaches while circumventing their disadvantages.


\section{Methodology} \label{sec:method}

\figu{}~\ref{fig:work-flow} shows the detailed workflow of \tool{}.
To test a \rstservice{}, \tool{} first takes its OpenAPI specifications as input to build the \atmab{} --- a novel representation of the relations among \rstapi{}s and the schemas (\S{}~\ref{sec:atmbuild}).
After building the initial \atmab{}, \tool{} uses it to generate call sequences and replace the API calls in the call sequences with actual requests (\S{}~\ref{sec:seqgen}).
Then the generated test cases are fed to the target \rstservice{} and \tool{} shall collect the responses.
By analyzing the collected responses, \tool{} reports the detected failures for bug analysis.
Moreover, \tool{} also uses the responses to refine the \atmab{} by adding missing edges and removing infeasible edges (\S{}~\ref{sec:atmupdate}).
The refined \atmab{} is then used for generating more test sequences.
This marks the end of one iteration and \tool{} will keep testing the target \rstservice{} and refining the \atmab{} until the time budget is reached.

\subsection{Initial \atmab{} Building}~\label{sec:atmbuild}
In \tool{}, we propose the concept of \atmab{} to encode the CRUD relations of operations, the relations of the schemas, and the data-flow among the schemas and operations.
The design of \atmab{} is based on the concept of \textit{property graph}~\cite{graph-paper}.
The definition of \textit{property graph} is as follows:
\begin{myDef} [Property Graph] \label{def:pg}
A \textit{property graph} is a directed, edge-labeled, attributed multigraph  $G=(V,E,\lambda,\mu)$ where $V$ is a set of nodes (or vertices), $E$ is a set of directed edges, $\lambda:E\to\Sigma$ is an edge labeling function assigning a label from the alphabet $\Sigma$ to each edge and $\mu:(V \cup E)\times K \to S$ is a function assigning key(from K)-value(from S) pairs of properties to the edges and nodes.
\end{myDef}
Given the definition of \textit{property graph}, \atmab{} is defined as follows:

\begin{myDef} [\atm] \label{def:atm}
A \atmab{} is a mixed,edge-labeled, attributed multigraph $G=(V,E,\lambda,\mu)$ with:
\begin{itemize}
    \item $V = V_{schema} \cup V_{operation}$
    \item $E = E_{os} \cup E_{so} \cup E_{ss} \cup E_{oo}$
    \item $\lambda = \lambda_{so} \cup \lambda_{ss}$
    \item $\mu = \mu_{schema} \cup \mu_{operation}$
\end{itemize}
where $V_{schema}$ is a set of schema nodes, $V_{operation}$ is a set of operation nodes,
$E_{os}$ is a set of directed edges pointing from a node in $V_{operation}$ to a node in $V_{schema}$,
$E_{so}$ is a set of directed edges each of pointing from a node in $V_{schema}$ to a node in $V_{operation}$,
$E_{ss}$ is a set of undirected edges connecting two nodes in $V_{schema}$,
$E_{oo}$ is a set of undirected edges connecting two nodes in $V_{operation}$,
$\lambda_{so}$ is a set of labeling functions to label edges in $E_{so}$,
$\lambda_{ss}$ is a set of labeling functions to label edges in $E_{ss}$,
$\mu_{schema}$ is a set of functions to assign properties to nodes in $V_{schema}$, and $\mu_{operation}$ is a set of functions to assign properties to nodes in $V_{operation}$
\end{myDef}

Note that \atmab{} is not exactly a \textit{property graph} because it is a mixed graph while the latter is a directed graph and this is the only difference.
In the following of this paper, for simplicity, we use $o_{x}$ to represent an element of $V_{operation}$, $s_{x}$ to represent an element of $V_{schema}$, $e_{os}$ to represent an element of $E_{os}$ and so on.

\begin{figure}[t]
     \centering
     \begin{subfigure}[b]{0.9\linewidth}
         \centering
         \includegraphics[width=\textwidth]{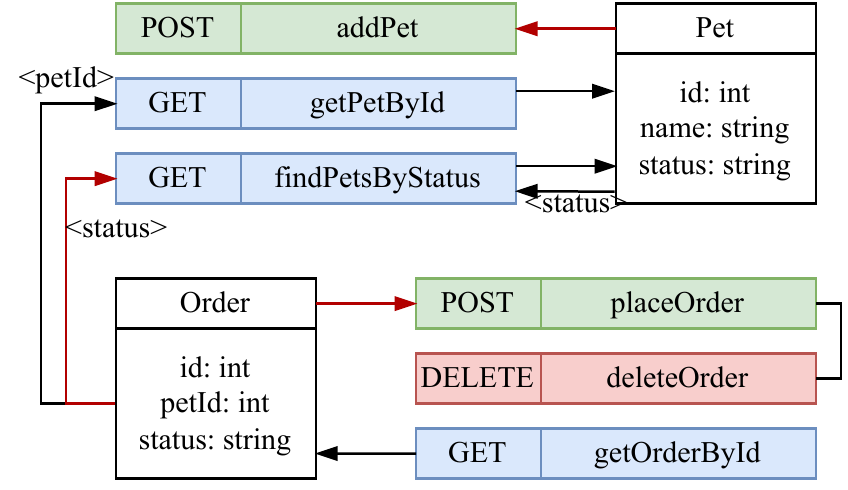}
         \caption{The Initial \atmab{}}
         \label{fig:atm-initial}
     \end{subfigure}
     \hfill
     \begin{subfigure}[b]{0.9\linewidth}
         \centering
         \includegraphics[width=\textwidth]{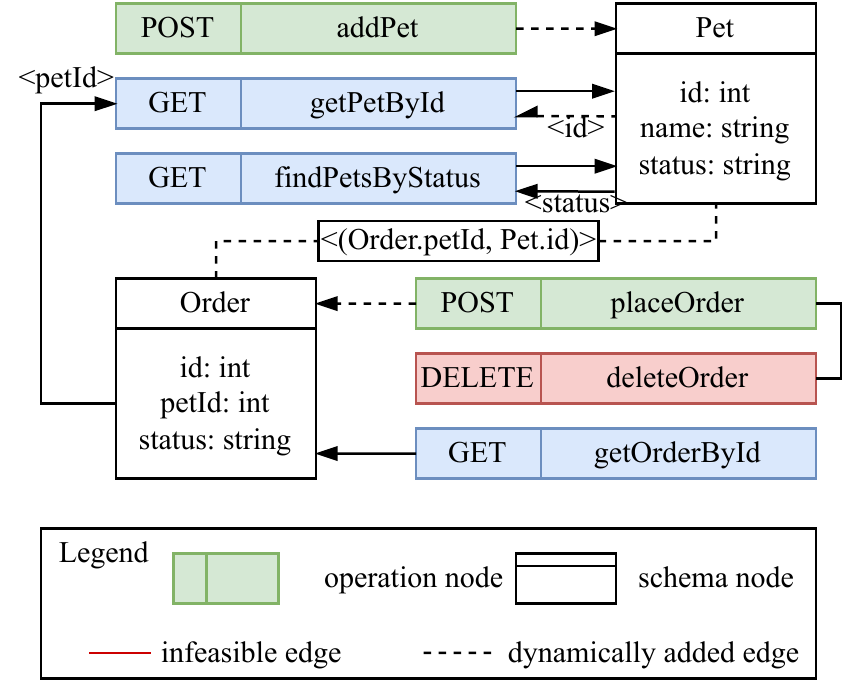}
         \caption{The Refined \atmab{}}
         \label{fig:atm-fixed}
     \end{subfigure}
     \caption[Caption for LOF]{The \atm{}s (\atmab{}s) for the Petstore running example.}
	\label{fig:dpe}
\end{figure}

\header{Design \& Rationale.} Here we explain the details and rationale for the design of \atmab{}.
\ding{182}
The usage of $V_{schema}$ and $V_{operation}$ are straight forward, we need them to represent the schemas and operations.
\ding{183}
As for edges, the edges in $E_{os}$ and $E_{so}$ are directed to represent whether an operation produces or consumes an object of a schema.
For example, the edge $e_{o1s1} = (o_1, s_1)$ means operation $o_1$ returns an object of schema $s_1$ as its response.
Correspondingly, the edge $e_{s1o1} = (s_1, o_1)$ means operation $o_1$ requires an object or at least one property of the object of schema $s_1$ in its parameter(s).
The edges in $E_{ss}$ represent the equivalence relation between two properties from two different schemas.
In the running example, the property \texttt{id} in the Pet schema is referring to the same thing as the property \texttt{petId} in the Order schema.
We denote the edge connecting these two schemas as $e_{s_{pet}s_{order}} = \{s_{pet}, s_{order}\}$.
The edges in $E_{oo}$ are used to connect two operations if they are under the same API endpoint.
In the running example, the operation \texttt{getOrderById} and the operation \texttt{deleteOrder} are connected by this type of edge.
\ding{184}
Only the edges in $E_{so}$ and $E_{ss}$ are labeled by functions in $\lambda$.
The labels for $E_{so}$ are vectors of property names showing which exact properties of a schema are used as parameters for an operation.
Note that the empty vectors are ignored in \figu{}~\ref{fig:dpe}.
The labels for $E_{ss}$ are vectors of tuples, where each tuple is pair of properties with equivalence relations from two different schemas.
The edges in $E_{oo}$ requires no labelling since they can only indicate two operations belong to the same API endpoint.
The labelling for $E_{os}$ is ignored for two reasons.
First, labelling edges is an error-prone process, especially when the OpenAPI specification is poorly written, the more edges we label, the more errors we may introduce.
Second, comparing with the edges in $E_{so}$, the edges in $E_{os}$ are less important for successfully requesting API operations because the former ones directly affect the input parameters of the operations.
\ding{185}
Naturally, the functions to assign properties to the nodes ($\mu$) correspond to the process of parsing the OpenAPI specification and extracting data from it.

With the OpenAPI specifications, \tool{} can build the initial \atmab{} consisting of nodes and edges together with their properties or labels according to \defi{}~\ref{def:atm}.
This initial \atmab{}, like shown in \figu{}~\ref{fig:atm-initial}, may contain false edges or miss some edges and \tool{} will refine it with dynamically collected feedback later on.

\header{\atmab{} vs ODG.}
Here we compare the \atmab{} model with the ODG model.
The first difference is that ODG has only the operation nodes and their producer-consumer dependencies as edges while \atmab{} has more types of nodes and edges.
The second difference is that \atmab{} allows nodes and edges to have properties and labels.
Both of the differences indicate that \atmab{} can describe more details of the \rstservice{}s.
The additional details captured by \atmab{} can help to generate longer call sequences and provide the information needed for dynamic self-updating.
For example, for the \atmab{} shown in \figu{}~\ref{fig:atm-fixed}, the connection between the \texttt{Pet} and \texttt{Order} schemas indicates that \texttt{Order.petId} refers to the same thing as \texttt{Pet.id}.
With this information, we can infer that the \texttt{Order.petId} can be used to query \texttt{getPetById} and therefore generate the call sequence (\texttt{placeOrder}, \texttt{getPetById}, \texttt{findPetsByStatus}), which cannot be generated only with the dependencies between operations in an ODG.

\subsection{Test Case Generation}~\label{sec:seqgen}
Here we introduce how \tool{} uses \atmab{} for test case generation.
In general, this requires two steps:
first, \tool{} traverses the \atmab{} and aggregates visited operations to form  call sequences (\textit{Call Sequence Generation});
second, \tool{} generates concrete inputs for the APIs in the sequences to build test cases (\textit{Operation Parameter Generation}).

\begin{algorithm}[t]
\setstretch{0.9}
\small
\SetKw{Continue}{\textbf{continue}}
\SetKw{Ret}{\textbf{return}}
\SetKw{And}{\textbf{and}}
\SetKw{Not}{\textbf{not}}
\SetKw{Is}{\textbf{is}}
\SetKw{Or}{\textbf{or}}
\SetKw{Eqs}{\textbf{equals}}
\SetKw{Fori}{\textbf{for}}
\SetKw{inv}{\textbf{in}}
\SetKwProg{Def}{def}{:}{}

\Def{call\_sequence\_generation($V, E, \lambda, \mu$)} {
$ \mathcal{S} \gets \emptyset$\;
\For{$s \in V_{schema}$}{
    $\mathcal{S} \gets \mathcal{S} \cup visit(s, \emptyset, V_{schema},  \emptyset)$\;
}
$ \mathcal{S} \gets crud\_filter(\mathcal{S})$\;
\Ret $\mathcal{S}$\;\label{line:ret}
}

$ $

\Def{visit($s$, $V_{visited\_schema}$, $V_{schema}$, $\mathcal{S}$)}{\label{line:startrt}
  $O_{out} \gets \{o \in V_{operation}\,|\, (o, s) \in E_{os}\}$\;
  $O_{in} \gets \{o \in V_{operation}\,|\, (s, o) \in E_{so}\}$\;
  \If{$O_{in} = \emptyset \;\lor\; O_{out} = \emptyset$}{
    \Ret $\mathcal{S}$\;
  }
  \If{$\{e_{ss} \in E_{ss}\,|\, e_{ss} = \{s, s'\}\,\land\, s' \neq s,\, s' \in V_{schema}\} = \emptyset$}{
    \Ret $\mathcal{S}$\;
  }
  $V_{visited\_schema} \gets V_{visited\_schema} \cup \{s\}$\;
  $\mathcal{S}_{new} = O_{out} \times O_{in}$\; \label{line:cproduct}
  $\mathcal{S} \gets concat(\mathcal{S}, \mathcal{S}_{new})$\;\label{line:cconcat}
  \For{$s' \in \{s' \in V_{schema}\,|\, (\exists\, e_{ss'} \in E_{ss})[e_{ss'} = \{s, s'\} \,\land\, s \neq s'] \,\land\, s' \notin V_{visited\_schema}\}$}{
    $\mathcal{S} \gets \mathcal{S} \cup visit(s', V_{visited\_schema}, V_{schema},\mathcal{S})$\;
  }\label{line:traverse_neighbor}
  \Ret $\mathcal{S}$\;
}\label{line:endrt}

$ $

\Def{concat($\mathcal{S}$, $\mathcal{S}_{new}$)}{\label{line:startcon}
  \For{$\mathbb{S} = (o_{1},...,o_{n}) \in \mathcal{S}$}{
    \For{$\mathbb{S}'= (o_{out},o_{in}) \in \mathcal{S}_{new}$}{
      \If{$o_{n} = o_{out}$}{
         $\mathcal{S} \gets \mathcal{S} \cup \{(o_{1},...,o_{n},o_{in})\}$\;
      }
      \If{$o_{in} = o_{1}$}{
        $\mathcal{S} \gets \mathcal{S} \cup \{(o_{out},o_{1},...,o_{n})\}$\;
      }
    }
  }
  \Ret $\mathcal{S}$\;
}\label{line:endcon}

$ $

\Def{crud\_filter($\mathcal{S}$)} {\label{line:startfilter}
\For{$\mathbb{S} \in \mathcal{S}$}{
    \If{\Not crud\_valid($\mathbb{S}$)}{
        $\mathcal{S} \gets \mathcal{S} - \{\mathbb{S}\}$\;
    }
}
\Ret $\mathcal{S}$\;\label{line:endfilter}
}

\caption{Call Sequence Generation}
\label{algo:seqgen}
\end{algorithm}

\header{Call Sequence Generation.}
\algo{}~\ref{algo:seqgen} shows how \tool{} generates call sequences where $\mathcal{S}$ is a set of call sequences and $\mathbb{S} = (o,o_{1},...,o_{n})$ is a single call sequence made up of a vector of operations.
The function \emph{call\_sequence\_generation} is the start of the whole process.

In general, the idea of this algorithm is to visit every schema in the \atmab{} and collect the operations related to the schema to build up call sequences.
While accessing a schema, we can recursively traverse other schemas which are connected to the current one, collect the related operations to build new call sequences and concatenate the new call sequences and the existing call sequences to build longer sequences.
After we have generated the call sequences, we further apply a \emph{crud\_filter} to filter out the call sequences violating the CRUD rules.
A call sequence is said to violate the CRUD rule if it encounters any of the two situations: for the same schema, a delete operation appears on the sequence before another operation; for the same schema, any of the read, update or delete operation appears on the sequence before the first create operation.

In particular, when we visit a schema $s$ (line~\ref{line:startrt}), we first need to identify two sets of operations: the operations which produce schema $s$ in their responses ($O_{out}$) and the operations which consume schema $s$ or its properties as their input parameters ($O_{in}$).
With $O_{out}$ and $O_{in}$, we can use their Cartesian product to build a set of call sequences ($\mathcal{S}_{new}$) with the length of two (line~\ref{line:cproduct}).
Then, if we already have a set of call sequences ($\mathcal{S}$), we can try to concatenate them and the new call sequences to build longer sequences (line~\ref{line:cconcat}).
Two sequences can be concatenated if the last operation of one of them is the same as the first operation of the other one (line~\ref{line:startcon}).
After that, if the current schema node is connected to another schema node which has not been visited in the current iteration (line~\ref{line:traverse_neighbor}), we will visit that schema and repeat the same process until we run out of schema nodes.
Note that each schema node is visited only once per iteration to avoid infinite loops.

In the running example, with the \atmab{} in \figu{}~\ref{fig:atm-fixed}, assume we are now visiting the \texttt{Pet} node and the \texttt{Order} node is not visited yet.
For the \texttt{Pet} node, its $O_{out}$ is \{\texttt{addPet}, \texttt{getPetById}, \texttt{findPetsByStatus}\} and its $O_{in}$ is \{ \texttt{getPetById}, \texttt{findPetsByStatus}\}.
The Cartesian product of $O_{out}$ and $O_{in}$ can generate six new call sequences including (\texttt{getPetById}, \texttt{findPetsByStatus}) etc. (line~\ref{line:cproduct}).
Since we do not have any existing call sequences so far, we do not need to perform call sequence concatenation (line~\ref{line:cconcat}).
Then, because the \texttt{Order} node is connected to the \texttt{Pet} node and it has not been visited, we will visit it and generate new call sequences similarly (line~\ref{line:traverse_neighbor}).
One of the newly generated call sequences for the \texttt{Order} node is (\texttt{placeOrder}, \texttt{getPetById}) (line~\ref{line:cproduct}).
Now, since we already have generated some sequences when we visit the \texttt{Pet} node, we can try to concatenate them with the newly generated call sequences for the \texttt{Order} node (line~\ref{line:cconcat}).
In this example, we can generate the call sequence (\texttt{placeOrder}, \texttt{getPetById}, \texttt{findPetsByStatus}).
This demonstrates how \tool{} generates call sequences.

Note that, in some cases, there are standalone operations that are not connected to any schemas.
For clarity, we omit handling these cases in the algorithm.
But in our implementation, the standalone operations are included as single-element call sequences.

\header{Operation Parameter Generation.}
The generated call sequences cannot be used directly for testing since they are just sequences of operations without concrete parameter values.
For a given operation, if its input parameters are \textit{not} from the responses of other operations on the same test sequences, they are decided as follows:
if the operation has been called correctly before, \tool{} assigns a high chance of using the parameters of the last successful run, else if the OpenAPI specifications have specified the valid ranges of values for the operation, \tool{} has a high chance of selecting values from the valid ranges, otherwise, \tool{} will just use random values.
Given an object schema, MOREST recursively traverses the schema to generate values for the object attributes according to their parameter types. If the parameter is of a basic type (e.g., string and int), we will concretize the value accordingly. If the parameter is an object or a list, we will jump in and recursively deduce the basic parameter type.

An important detail worth discussing is that the parameters for a sequence of operations is not generated all at once.
Instead, these parameters are generated on the fly.
Because if the input parameter for an operation is based on the response of a previous operation, we will need to wait for the previous operation to finish execution to get the needed parameter values.


\subsection{Dynamic \atmab{} Updating}~\label{sec:atmupdate}
The initial \atmab{} generated with the OpenAPI specifications may contain errors.
For example, the red lines in \figu{}~\ref{fig:atm-initial} are infeasible edges, and the dashed lines in \figu{}~\ref{fig:atm-fixed} are the missing edges for the initial \atmab{}.
This is often due to ambiguities in the specifications.
For example, although a human can quickly recognize that the \texttt{petId} property of \texttt{Order} refers to the same thing as the \texttt{id} property of \texttt{Pet}, it is hard to use rules and heuristics to reconstruct this relation.
To address this problem, \tool{} dynamically fixes the \atmab{} with execution feedback.

\header{Edge Addition}
In \tool{}, new edges are added to a \atmab{} in three scenarios:
\ding{182} The most straightforward case is where the response value of an API aligns with a certain schema but it is not documented in the specification.
For example, in \figu{}~\ref{fig:api-definition}, the response of the \texttt{addPet} operation is not documented.
However, after \texttt{addPet} is corrected during the testing, it will respond with the newly created \texttt{Pet} object.
\tool{} will notice that the properties of the object returned by \texttt{addPet} matches the properties in the \texttt{Pet} schema.
Then \tool{} can add the edge (\texttt{addPet},\texttt{Pet}) into $E_{os}$.
\ding{183} The edges of $E_{ss}$ are added after the collection of execution feedback.
For two schema nodes $s_1$ and $s_2$ if both of them are related to an operation node $o$, \tool{} makes the assumption that $s_1$ and $s_2$ share at least one property in common but \tool{} cannot decide which property is shared in the \atmab{} building stage.
During the execution of the test cases, \tool{} may encounter some sequences involving both the objects of $s_1$ and the objects of $s_2$.
With the responses of these sequences, \tool{} can compare the concrete values of properties between the two types of objects.
As far as the value of a property from an object of $s_1$ can match the value of multiple properties from an object of $s_2$, the relation between $s_1$ and $s_2$ remains undecided.
Until we find a property whose value only matches with one property from the other set of objects, we can tell that these two schemas share this common property.
For example, both \texttt{Order} and \texttt{Pet} connect with \texttt{getPetById} (\figu{}~\ref{fig:atm-initial}).
\tool{} can generate multiple test cases with the sequence of (\texttt{getOrderById}, \texttt{getPetById}).
Suppose these test cases yield the following two pairs of orders and pets:
(\texttt{\{id: 1, petId: 1, status: "succ"\}, \{id: 1, name: "cat", status: "sold"\}}) and (\texttt{\{id: 2, petId: 4, status: "succ"\}, \{id: 4, name: "dog", status: "sold"\}}).
After the generation of the first pair of order and pet, \tool{} cannot infer the relation since the \texttt{id} of the pet equals to both \texttt{id} and \texttt{petId} of the order.
\tool{} can only infer that \texttt{Pet.id} is equal to either \texttt{Order.id} or \texttt{Order.petId}.
Nevertheless, after the second pair is generated, \tool{} can draw the conclusion of \texttt{Pet.id} equals to {Order.petId}.
\ding{184} The last scenario is where the edges belonging to $E_{so}$ and $E_{os}$ are added based on the inferences used for building $E_{ss}$.
For example, after \tool{} learns the fact that \texttt{Pet.id} equals to {Order.petId}, it propagates this information to other nodes, say \texttt{getPetById}.
Then, \tool{} can link the \texttt{petId} parameter needed by \texttt{getPetById} with the property \texttt{Pet.id} despite that they have different names.
As a result, \tool{} can add the edge (\texttt{Pet}, \texttt{getPetById}) to $E_{so}$ as shown in \figu{}~\ref{fig:atm-fixed}.

\header{Edge Deletion}
For an operation, if its inputs are from other operations on the same call sequence (i.e. not generated randomly) and it fails to execute correctly after $\Theta$ tries, then the edge pointing from the respective schema to this operation is temporarily recognized as infeasible.
Empirically, we find that increasing the value of $\Theta$ can help to reduce the number of falsely deleted edges but the overall performance is not affected too much since we are wasting more time on the truly infeasible edges and even if a benign edge is deleted, \tool{} always has the chance of adding it back.

\section{Implementation \& Evaluation}\label{sec:evaluation-setup}

\begin{table}[]
\caption{Open source \rstservice{}s}
\label{tab:testing-subjects}
\setlength\tabcolsep{1pt}
\resizebox{\linewidth}{!}{
\begin{tabular}{c||cccccc}
\hline
Subjects & LoC    & Language & Operation & Endpoint   & Source    \\ \hline
Petstore~\cite{swagger-petstore}            & 989    & Java & 20 & 18    & OpenAPI   \\ 
SpreeCommerce~\cite{evomaster}      & 42,385 & Ruby & 35 & 31 & EMB       \\ 
Bitbucket~\cite{bitbucket-home}        & 46,321     & Java  & 32 & 23    & Bitbucket \\ 
LanguageTool~\cite{evomaster}       & 4,760  & Java & 5 & 5   & EMB       \\ 
FeatureService~\cite{evomaster}     & 1,525  & Java    & 18 & 11  & EMB       \\ 
Magento~\cite{evomaster}     & 315,120  & PHP & 47 & 39      & EMB       \\ \hline
\end{tabular}
}
\end{table}

We have implemented \tool based on Python 3.9.0 with 11,659 lines of code and conducted experiments to evaluate the performance of \tool{}.
We aim to answer the following research questions with the evaluation:
\\
\listitem{RQ1 (Coverage)} How is the code coverage and operation exploration capability of \tool{}?

\listitem{RQ2 (Bug Detection)} How is the bug detection capability of \tool{}?

\listitem{RQ3 (Ablation Study)} How do \atmab{} guidance and dynamic \atmab{} updating affect the performance of \tool{} separately?



\subsection{Experiment Setup}
\header{Evaluation Datasets.}
We built the evaluation benchmark with six open source \rstservice{}s shown in \tabl{}~\ref{tab:testing-subjects}.
In this benchmark, four services were selected from the \evomaster{} Benchmark (EMB)~\cite{evomaster}~\footnote{We only use the open source services available in EMB up to the time of writing this paper.}.
In addition, Petstore (the official demo of OpenAPI specification) and Bitbucket (a popular online version control system with complicated business scenarios) were also included.

From \tabl{}~\ref{tab:testing-subjects}, we can see that the target \rstservice{}s are diverse in sizes, features and are implemented with different programming languages.
Bitbucket is not strictly open source since it is a commercial product.
However, we have access to its released .jar file to collect line coverage and conduct detailed bug analysis.

\header{Evaluation Baselines.}
We use three state-of-the-art blackbox \rstservice{} testing techniques as baselines to study the performance of \tool{}. 
\begin{enumerate}[leftmargin=*]
\item \listitem{\evomasterbb{}}:
\evomaster{}~\cite{evomaster} is a search-based whitebox technique.
In our evaluation, we disabled its code instrumentation (only works for Java programs) and database monitoring modules to turn it into blackbox mode (named as \evomasterbb{}).
It uses various heuristics to generate call sequences according to the OpenAPI specifications.
Thus, \evomasterbb{} represents the heuristic-based approach.


\item \listitem{\restgen{}}:
\restgen{}~\cite{resttestgen} is a blackbox technique which generates Operation Dependency Graphs (ODGs) from OpenAPI specifications to guide the call sequence generation.
\restgen{} represents the top-down approach.


\item \listitem{\restler{}}:
\restler{} is a blackbox technique which build call sequences by appending new API operations according to execution feedback.
\restler{} represents the bottom-up approach.

\end{enumerate}

\header{Evaluation Criteria.}
\label{sec:failure-def}
We use three criteria to evaluate \tool{} and the baselines.

\begin{enumerate}[leftmargin=*]
    \item \listitem{Code Coverage}:
    Code coverage can reflect the exploration capability of the techniques.
    In the experiments, we use line coverage since it is the finest granularity we can get with our current tools (PHPCoverage~\cite{php-coverage} for PHP, CoverBand~\cite{ruby-coverage} for Ruby and JaCoCo~\cite{jacoco} for Java).
    
    \item \listitem{Successfully Requested Operations}:
    In HTTP, responses with status code 2xx are successful responses~\cite{statuscode}.
    We use the number of successfully requested operations (SROs) as a criterion because it can reflect whether a technique can generate valid requests to test deeper code logic of the \rstservice{} and valid requests are partially the results of correct call sequences.
    \item \listitem{Bugs}:
    The goal of testing is to expose bugs, so the number of detected bugs is a necessary criterion.
    In the context of \rstservice{}, a \emph{failure} is considered to happen when an operation returns 5xx status code and a \emph{bug} can be related to many failures.
    We manually classified the failures into unique bugs in the experiments according to response bodies, server logs, etc.
\end{enumerate}

\header{Evaluation Settings.}
For the open source \rstservice{}s, we hosted them on a local machine and ran each technique with three time budgets --- 1, 4 and 8 hour(s).
The purpose is to evaluate how the performance of these techniques change over time.
To the best of our knowledge, the time budget of 8 hours is the longest time budget used in \rstservice{} testing research.
After each round, we tear down and restore the environment (e.g., docker containers, self-hosted virtual machines) to guarantee the consistence of all \rstservice{}s. In addition, we repeated all experiments for 5 times to mitigate randomness and applied Mann-Whitney U test and $\hat{A}_{12}$~\cite{a12,mann-u-test} calculation for statistical tests.
Thus, we conducted a total number of 1,440, i.e., 6 projects * 6 settings * 8 (hour time budget for one round) * 5 repetitions, CPU hours of experiments.



\begin{figure*}[t]
     \centering
     \begin{subfigure}[b]{0.33\linewidth}
         \centering
         \includegraphics[width=\textwidth]{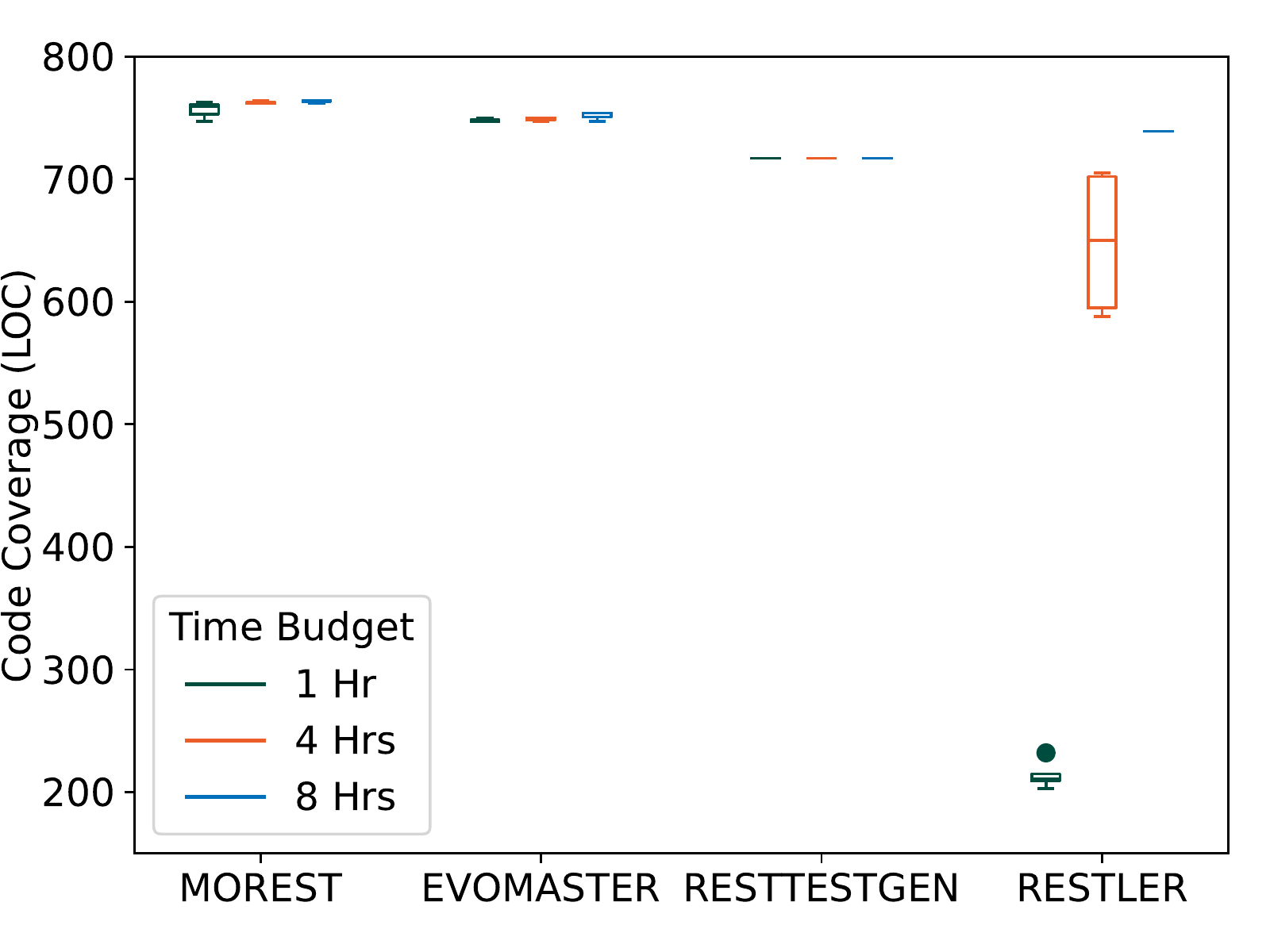}
         \caption{PetStore}
         \label{fig:code-coverage-petstore}
     \end{subfigure}
     \hfill
     \begin{subfigure}[b]{0.33\linewidth}
         \centering
         \includegraphics[width=\textwidth]{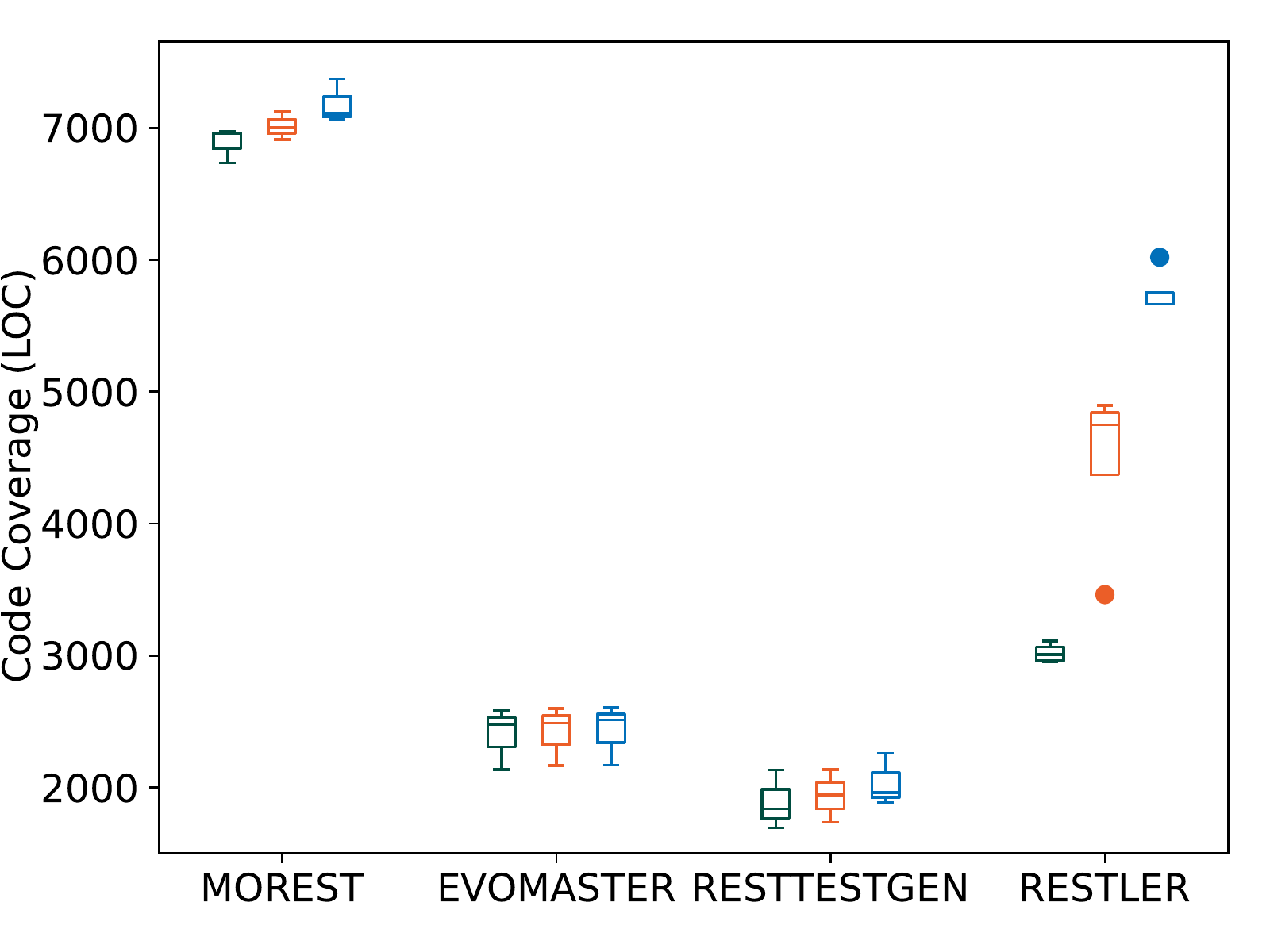}
         \caption{SpreeCommerce}
         \label{fig:code-coverage-spreecommerce}
     \end{subfigure}
     \hfill
     \begin{subfigure}[b]{0.33\linewidth}
         \centering
         \includegraphics[width=\textwidth]{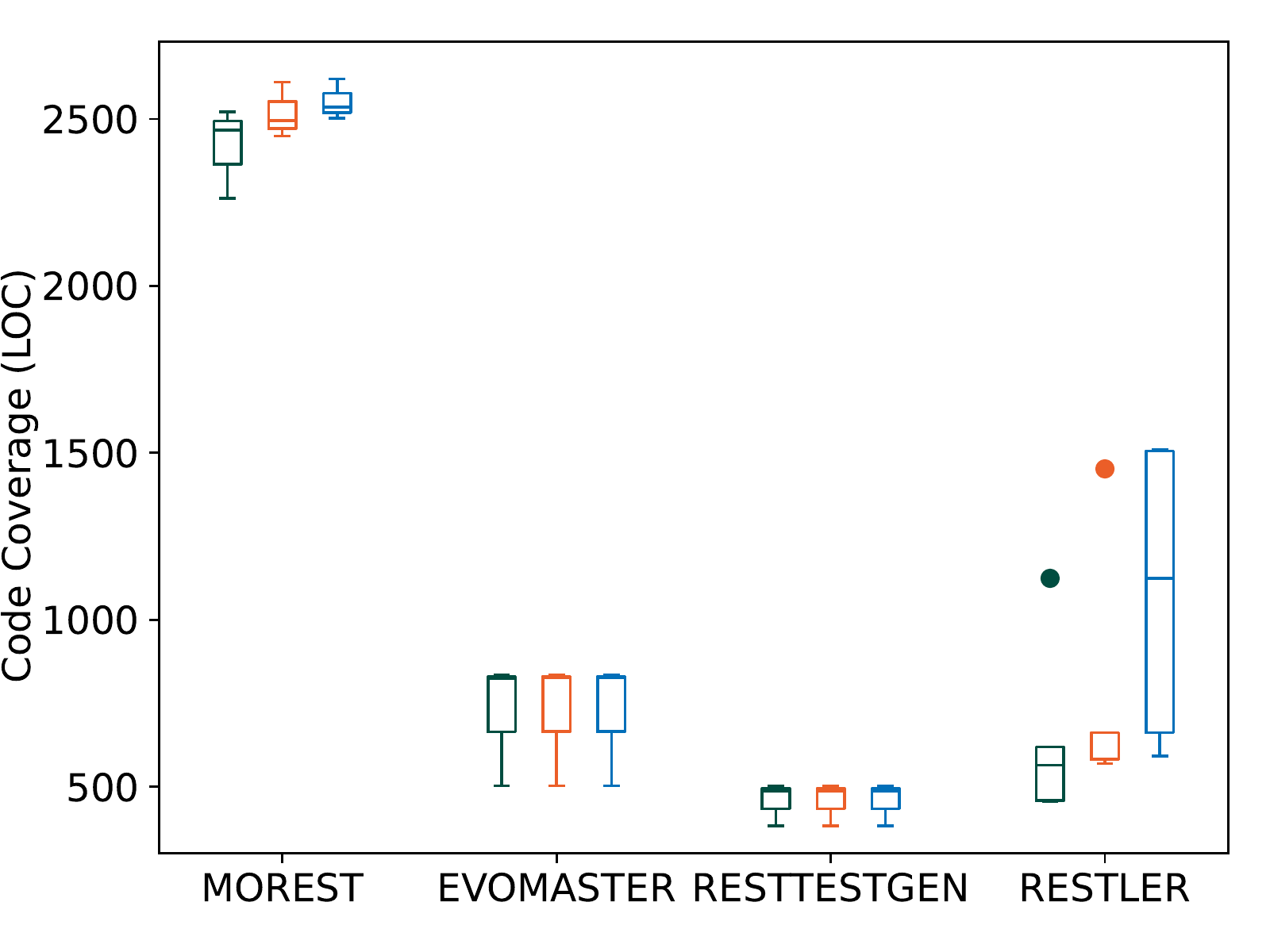}
         \caption{BitBucket}
         \label{fig:code-coverage-bitbucket}
     \end{subfigure}

      \begin{subfigure}[b]{0.33\linewidth}
         \centering
         \includegraphics[width=\textwidth]{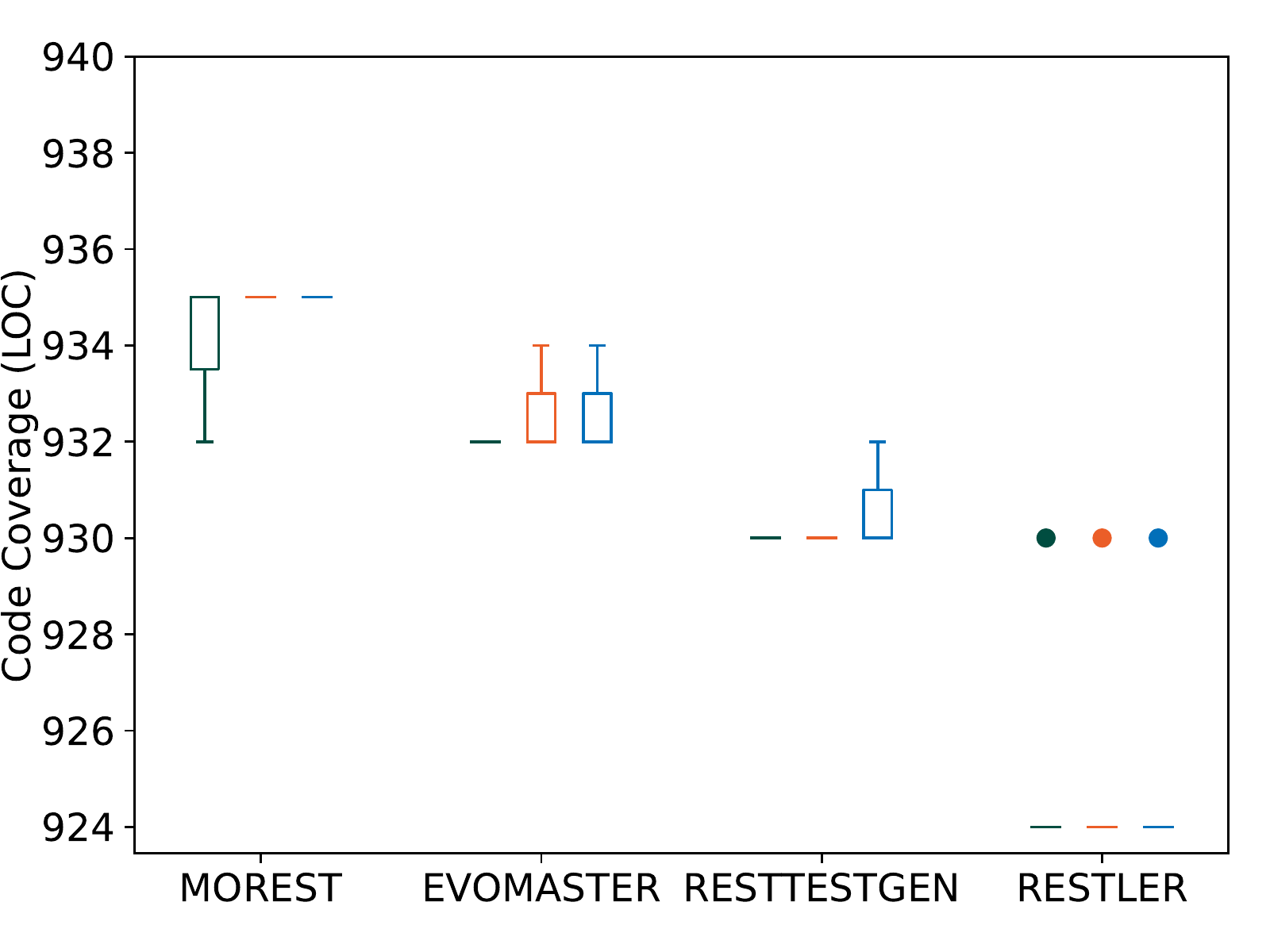}
         \caption{LanguageTool}
         \label{fig:code-coverage-languagetool}
     \end{subfigure}
     \hfill
     \begin{subfigure}[b]{0.33\linewidth}
         \centering
         \includegraphics[width=\textwidth]{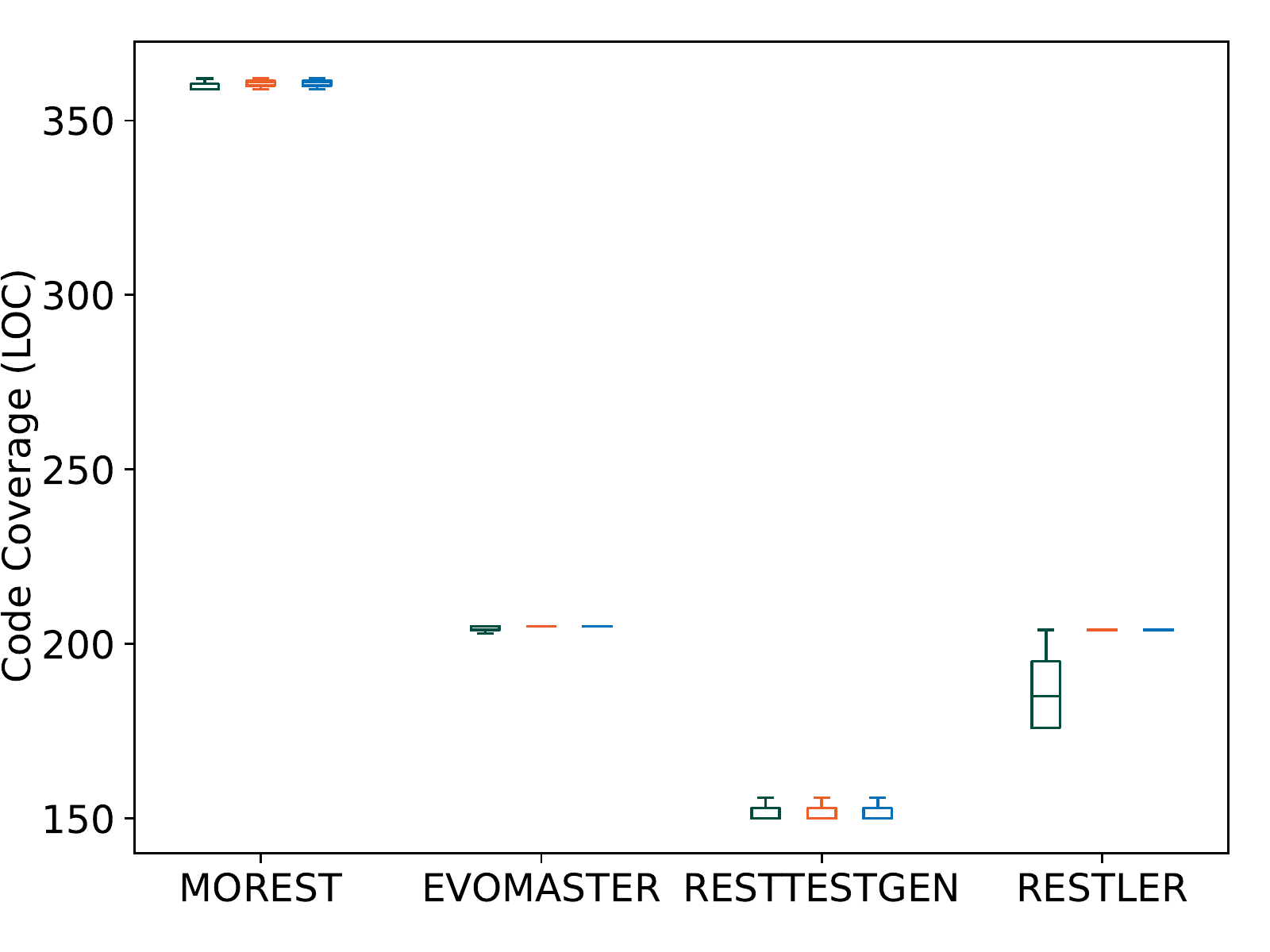}
         \caption{FeatureService}
         \label{fig:code-coverage-featureservice}
     \end{subfigure}
     \hfill
     \begin{subfigure}[b]{0.33\linewidth}
         \centering
         \includegraphics[width=\textwidth]{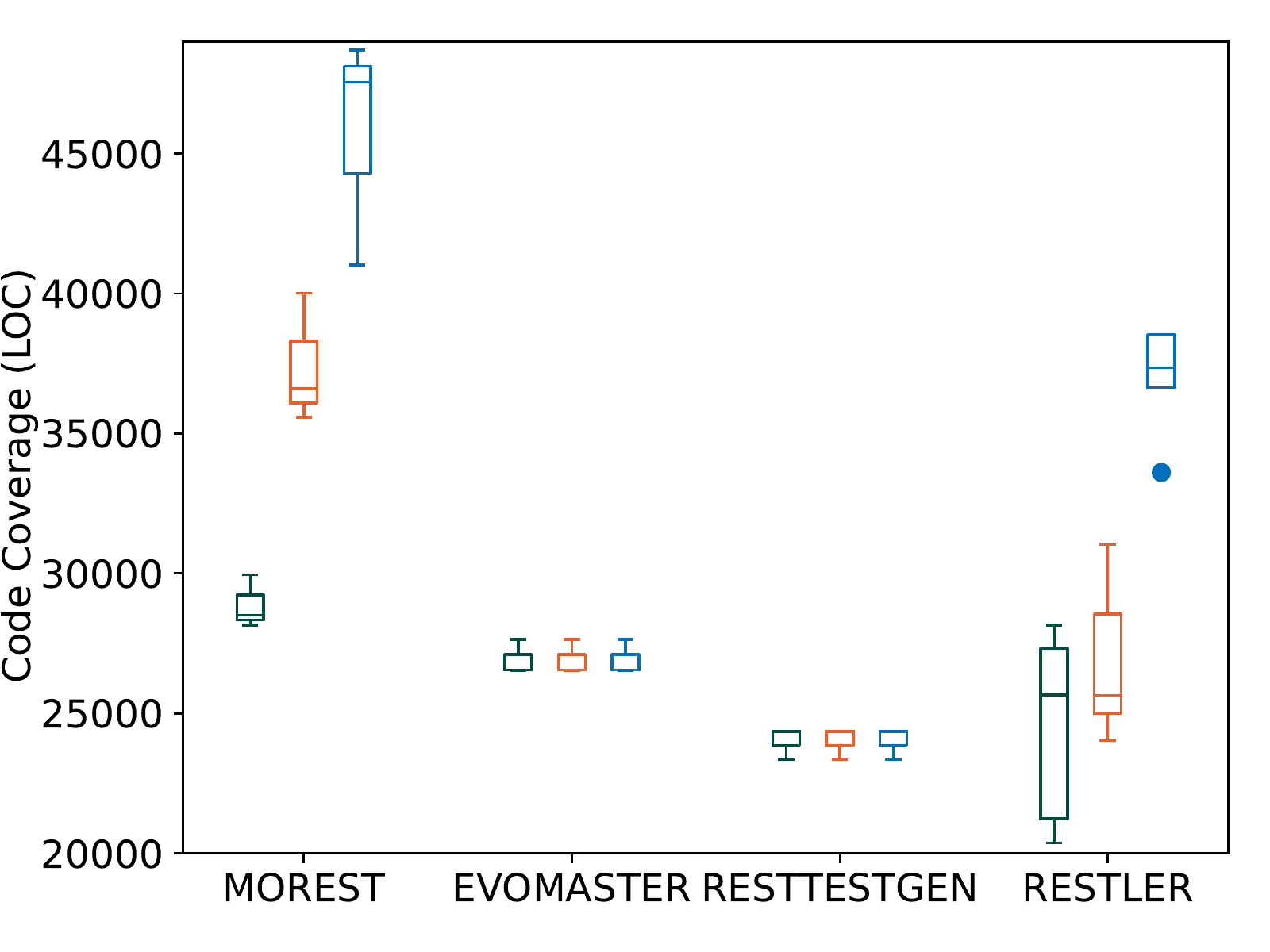}
         \caption{Magento}
         \label{fig:code-coverage-magento}
     \end{subfigure}
     \caption[Comparing with baselines in terms of code coverage ($\mu LOC$)]{Comparing with baselines in terms of code coverage ($\mu LOC$)}
	\label{fig:code-coverage}
\end{figure*}

\begin{table*}[h]
			\caption{Performance of \tool over \evomasterbb, \restgen, and \restler in terms of both the code coverage ($\mu LoC$) and successfully requested operations (SRO). We run this experiment 5 times (8 hours each time) and highlight statistically significant results in bold. (We calculate the average increased number by $\frac{(\#\ of \ \tool)- (\# \ of \ baseline)}{\# \ of \ baseline}$.)
		}
		
		\label{tab:coverage}
        \resizebox{\textwidth}{!}{
        \begin{tabular}{c||c|cc|cc|cc||c|cc|cc|cc}
            \hline

            \multicolumn{1}{c||}{\multirow{3}{*}{Subjects}} & \multicolumn{7}{c||}{ Average Code Coverage (LoC)} &
            \multicolumn{7}{c}{Average \# of Successfully Requested Operations (SRO)}
            \tabularnewline
            \cline{2-15}
             & \tool & \multicolumn{2}{c|}{\evomasterbb} & \multicolumn{2}{c|}{\restgen} & \multicolumn{2}{c||}{\restler} & \tool & \multicolumn{2}{c|}{\evomasterbb} & \multicolumn{2}{c|}{\restgen} & \multicolumn{2}{c}{\restler} \tabularnewline

             & $\mu LOC$ & $\mu LOC$ & $\hat{A}_{12}$  & $\mu LOC$& $\hat{A}_{12}$  & $\mu LOC$ & $\hat{A}_{12}$  &  \multicolumn{1}{|c|}{$\mu SRO$ } & $\mu SRO$ & $\hat{A}_{12}$  & $\mu SRO$ & $\hat{A}_{12}$ & $\mu SRO$ & $\hat{A}_{12}$  \tabularnewline
             
            \hline 
           Petstore  & \textbf{763.20} & 751.80 & 1.00   & 717.00 & 1.00    & 739.00 & 1.00   & \textbf{20.00}    & 14.00 & 1.00    & 13.00 & 1.00  & 18.00 & 1.00    \tabularnewline

            SpreeCommerce  & \textbf{7182.00} & 2428.80 & 1.00   & 2036.00 & 1.00   & 5753.80 & 1.00  & \textbf{18.40}   & 1.00 & 1.00   & 1.00 & 1.00   & 1.00 & 1.00   
            \tabularnewline

            Bitbucket  & \textbf{2552.60} & 721.80 & 1.00  & 457.00 & 1.00  & 1078.80 & 1.00  & \textbf{15.00}   & 2.00 & 1.00   & 2.00 & 1.00 & 1.00 & 1.00    \tabularnewline

            LanguageTool  & \textbf{935.00}  & 932.80 & 1.00  & 930.8 & 1.00   & 925.20 & 1.00 & 1.00 & 1.00  & 0.50 & 1.00 & 0.50  & 1.00 & 0.50     \tabularnewline
            
            FeatureService  & \textbf{360.00}   & 205.00 & 1.00   & 152.00 & 1.00 & 204.00 & 1.00  & \textbf{18.00}    & 6.00 & 1.00   & 4.40 & 1.00 & 9.00 & 1.00  \tabularnewline
            
            Magento  & \textbf{45759.00}   & 26912.00 & 1.00   & 24024.40 & 1.00 & 36933.00 & 1.00  & \textbf{18.60}   & 8.00 & 1.00   & 6.00 & 1.00  & 6.00 & 1.00   
            
            \tabularnewline

            \hline 
             Average Increased (\%)  & 0.00  & 80.12  & -  & 103.24 & -  & 26.16  & -  & 0.00  & 184.42 & -  & 232.45 & -  & 152.66  & - \tabularnewline
             \hline 
            \end{tabular}
            }
\end{table*}

\begin{figure}[t]
  \centering
      \begin{subfigure}[b]{0.52\linewidth}
    \centering
    \includegraphics[width=\textwidth]{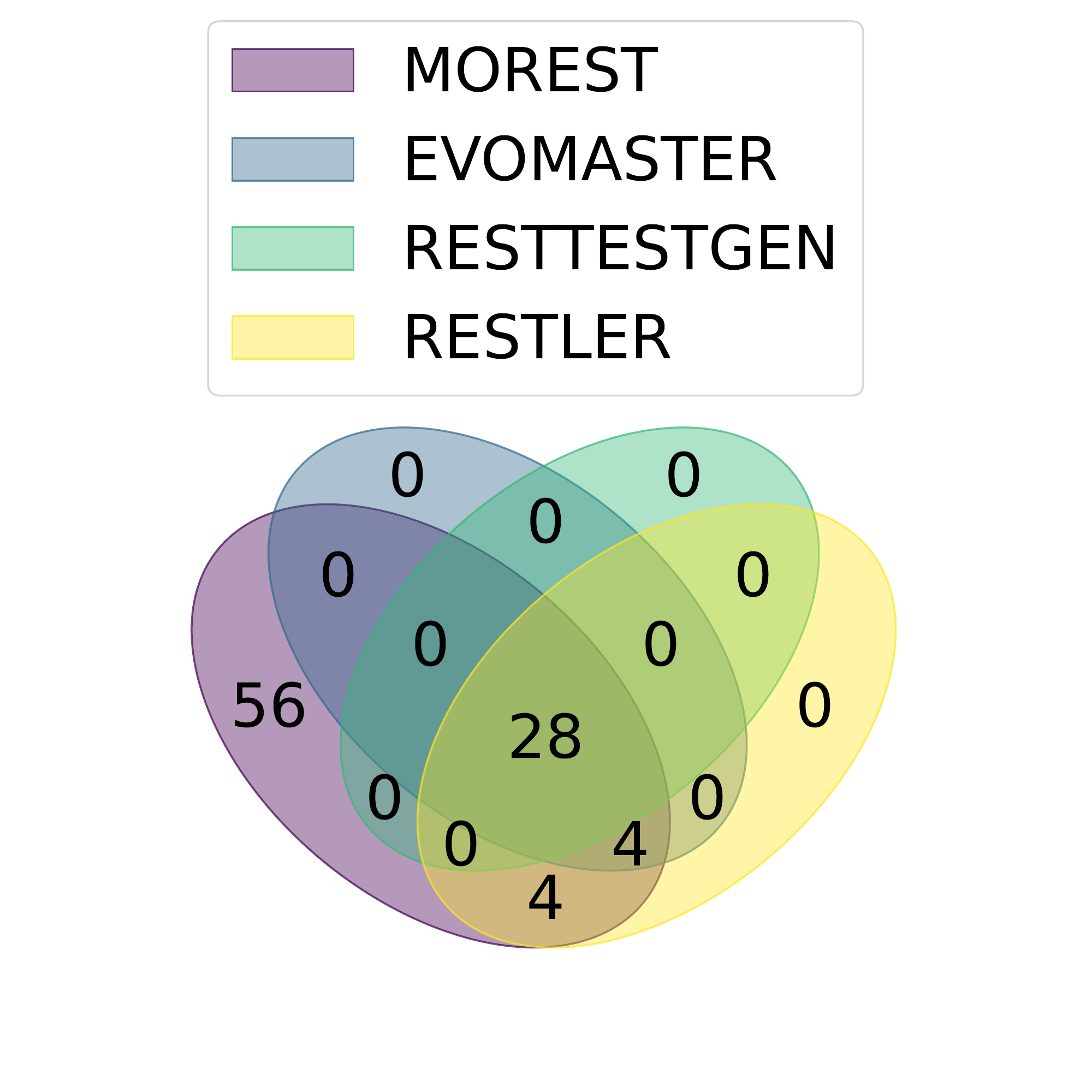}

             \caption{{\fontsize{6}{60}\selectfont Successfully Requested Operations (SRO)}}
                   \label{fig:venn-sro}

  \end{subfigure}
  \hfill
      \begin{subfigure}[b]{0.4\linewidth}
    \centering
    \includegraphics[width=\textwidth]{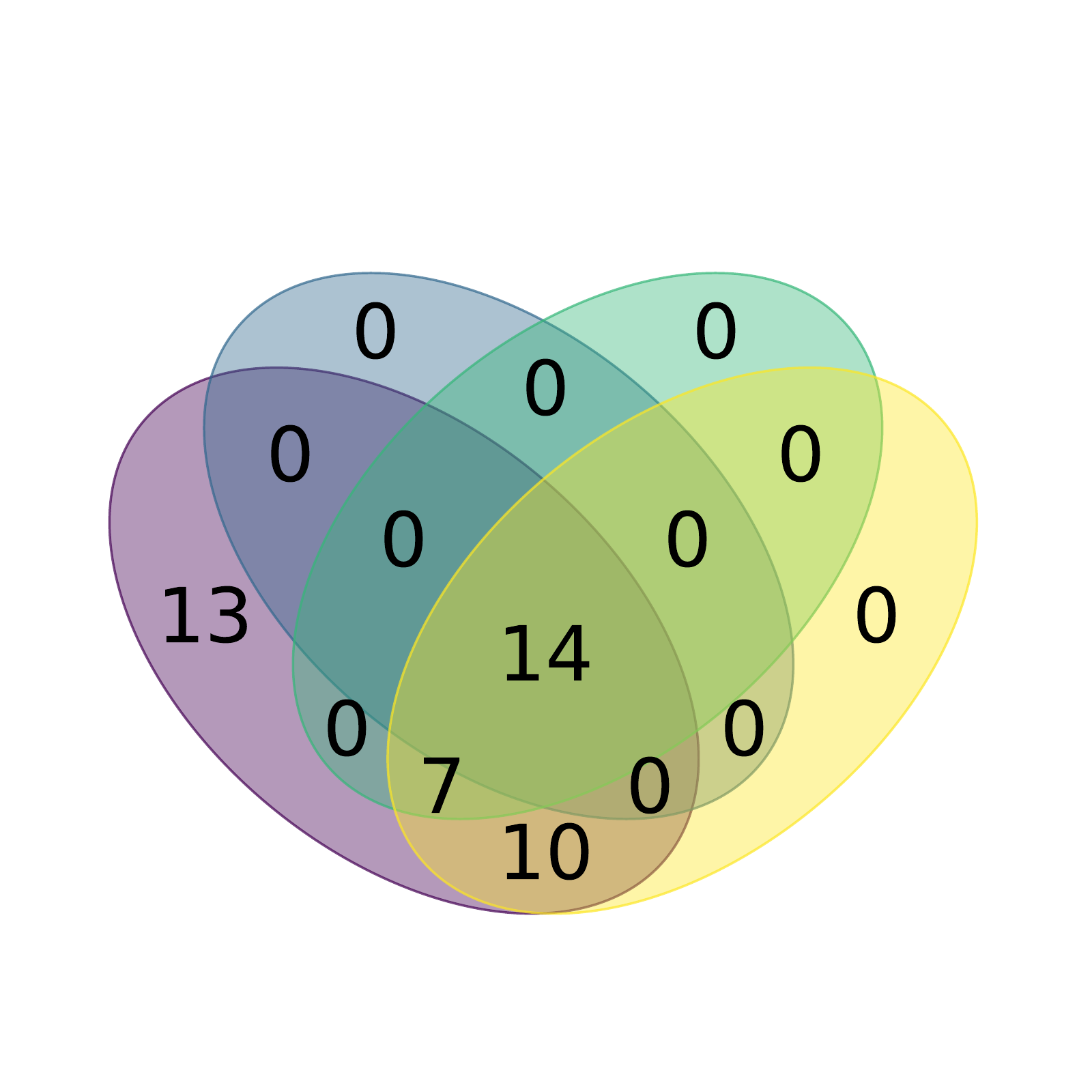}
             \caption{Bug detection}
                   \label{fig:venn-bug}

  \end{subfigure}
  
  \caption{Venn diagrams showing both successfully requested operations (SRO) and bug detection that  \tool{}, \evomasterbb{}, \restgen{} and \restler{} individually and together.}
  \label{fig:venn-coverage}
\end{figure}
\subsection{Coverage (RQ1)}\label{sec:evaluation-coverage}
To comprehensively compare different approaches, \evomasterbb, \restgen, and \restler are directly adopted from the prior work~\cite{evomaster,restler,resttestgen}. Besides, to avoid the side effect of service under test, we wrap six evaluation subjects into docker image~\cite{docker} and restore the corresponding image after each round. \tabl~\ref{tab:coverage} presents the statistical results of 
five runs. As suggested by the prior work~\cite{klees2018evaluating}, the Mann-Whitney U test (with the confidence threshold $\alpha = 0.05$) is adopted here. Overall, we summarize our findings as follows.

\figu{}~\ref{fig:code-coverage} and \tabl{}~\ref{tab:coverage} depict that~\footnote{Due to page limitation, we leave experiment result (code coverage and successfully requested operations), for 1 hour and 4 hours, in our website~\cite{morest}.} \tool achieves competitive performance regarding both code coverage and successfully requested operations, significantly outperforming search-based, and model-based approaches in 5/6 \rstservice{}s (bold numbers in \tabl~\ref{tab:coverage}). \figu{}~\ref{fig:venn-sro} demonstrates unique successfully requested operations by each tool. We can figure out that \tool covers the most operations by generating valid call sequences following producer-consumer dependencies. Therefore, existing experimental result demonstrates the exploration effectiveness of \tool regarding given coverage criteria (code coverage increased by 26.16\% - 103.24\%, successfully requested operations improved by 152.66\% - 232.45\%). On the other hand, it presents the robustness of \tool gained by \atmab{} guidance and \atmab{} updating fashion (experiment results are statistically significant).

Further, we conduct an in-depth analysis of the generated test sequences by all approaches to figure out why all approaches achieve the same number of successfully requested operations. Taking the LanguageTool as an example, we find that all approaches could not cover three operations getWords, addWord and deleteWord.  Two parameters, \textit{username} and \textit{APIKey}), in those operations are labeled as required parameters (i.e., \textit{username} and \textit{APIKey} must be filled in each request when testing those operations). However, we observe that the implementation of LanguageTool is different from 
descriptions in the OpenAPI specification. Specifically, in the implementation of LanguageTool, \textit{username} and \textit{APIKey} could not be filled simultaneously; otherwise, the server behaves unexpectedly and throws 5xx responses. 
As a consequence, we could imply that the gap between \rstapi specifications and actual implementation of \rstservice{} could affect the effectiveness of \rstapi testing.

\begin{table}[h]
		\caption{Comparing with baseline in terms of the number of detected bugs. 
		(Values in bold indicate statistically significant differences between \tool, \evomasterbb, \restgen, and \restler.)
		}
		\label{tab:bug}
        \resizebox{0.45\textwidth}{!}{
        \begin{tabular}{c||c|cc|cc|cc}
            \hline

            \multicolumn{1}{c||}{\multirow{3}{*}{Subjects}} & \multicolumn{7}{c}{ Average Detected Bugs (\#)}
            \tabularnewline
            \cline{2-8}
             & \tool & \multicolumn{2}{c|}{\evomasterbb} & \multicolumn{2}{c|}{\restgen} & \multicolumn{2}{c}{\restler}  \tabularnewline

             & $\mu \#$ & $\mu \#$ & $\hat{A}_{12}$  & $\mu \#$ & $\hat{A}_{12}$    & $\mu \#$  & $\hat{A}_{12}$    \tabularnewline
             
            \hline 
           Petstore  & \textbf{9.00}  & 3.80 & 1.00   & 0.00 & 1.00 & 8.00 & 1.00        \tabularnewline

            SpreeCommerce  & \textbf{3.00}   & 0.00 & 1.00   & 0.00 & 1.00   & 0.00 & 1.00 
            \tabularnewline

            Bitbucket  & \textbf{2.00}   & 0.00 & 1.00   & 0.00 & 1.00   & 0.00 & 1.00 
            \tabularnewline

            LanguageTool  &  3.00    & 3.00 & 0.50  & 3.00 & 0.50   & 3.00 & 0.50 
            \tabularnewline
            
            FeatureService  & \textbf{12.00}  & 6.00 & 1.00  & 11.60 & 0.67  & 10.00 & 1.00 
            \tabularnewline
            
            Magento  & \textbf{14.60} & 1.00 & 1.00  & 6.00 & 1.00  & 10.00 & 1.00
            \tabularnewline
            \hline
            Average Increased (\%)  & 0.00  & 215.94  & -  & 111.65 & -  & 40.64  & -  
            \tabularnewline
            
             
            \hline 
            \end{tabular}
            }
\end{table}

\subsection{Bug Detection (RQ2)}
Subsequently, we analyze the bug (defined in \sect~\ref{sec:failure-def}) discovering the capability of each approach. \tabl~\ref{tab:bug} presents the statistical results of the averaged bugs detection with running tools for 8 hours each round. Since one bug can be detected several times during testing, we only record the number of unique bugs for all approaches. Overall, we summarize our findings as follows.


First, in those baselines, \evomasterbb, \restler, and \restgen outperform each other in different subjects. We manually analyze the test sequences generated by baseline and find that \ding{182} \restler could sufficiently explore all call sequences theoretically, however, we have observed that massive invalid call sequences make \restler inefficient;\ding{183} \evomasterbb heuristically generates 
redundant call sequences to make it inefficient;\ding{184} \restgen is trapped by infeasible call sequences generated from the operation dependency graph, which limit its performance.

Besides, as shown in \tabl~\ref{tab:bug}, \tool detects the most bugs (bold numbers in \tabl~\ref{tab:bug}), and statistically outperforms other baselines on testing \rstapi services. Specifically, \figu{}~\ref{fig:venn-bug} indicates that \tool could detect unknown 13 bugs by existing approaches. It not only shows the comparative performance in bug finding but also the \tool's robustness.

Similarly to \sect~\ref{sec:evaluation-coverage}, we manually perform a comprehensive study in LanguageTool to figure out why all approaches find identical bugs. We find that due to the gap between \rstapi specification and the actual implementation of \rstservice{} (see detail description in \sect~\ref{sec:evaluation-coverage}), all approaches get 5xx responses by following the parameter's specifications (e.g., \textit{word\_size} is defined by the string type while implemented by the integer type) in the LanguageTool's specification. This indicates that \rstapi testing could help developers to identify the incompatibility between the \rstapi specifications and actual implementations.

\header{Case Study.}Two bugs, discovered by \tool, have been confirmed and fixed by the developers of BitBucket. We conduct a case study to find out how \tool trigers those bugs. Bitbucket is a Git-based source code repository hosting service owned by Atlassian. Bitbucket Server~\cite{bitbucket-server} is the self-host service provided by Atlassian that allows users to do unlimited API queries to the target server. During the testing of Bitbucket, we identified the following call sequence that triggers Internal Server Error with 500 status code.  In particular, (1) create a project at /rest/.../projects endpoint through POST operation. The project information can be further retrieved by GET request on the same endpoint. (2) create a repository in the project at /rest/.../\{projectKey\}/repos endpoint through POST operation, where the \{projectKey\} parameter is defined in the first step as a parameter. (3) A further GET query on /rest/.../repos/\{repositorySlug\}/ commits with parameters \{"path": "test\_string"\} triggers internal server error. When debugging mode is enabled, the bug message is printed out: "com.atlassian.bitbucket.\\scm.CommandFailedException". In summary, both a project and a repository should be created firstly to trigger this bug. With the \atmab{} guidance and dynamic \atmab{} updating, \tool could adaptively generate such call sequences.





  

\begin{figure}[t]
  \centering
      \begin{subfigure}[b]{\linewidth}
    \centering
    \includegraphics[width=\textwidth]{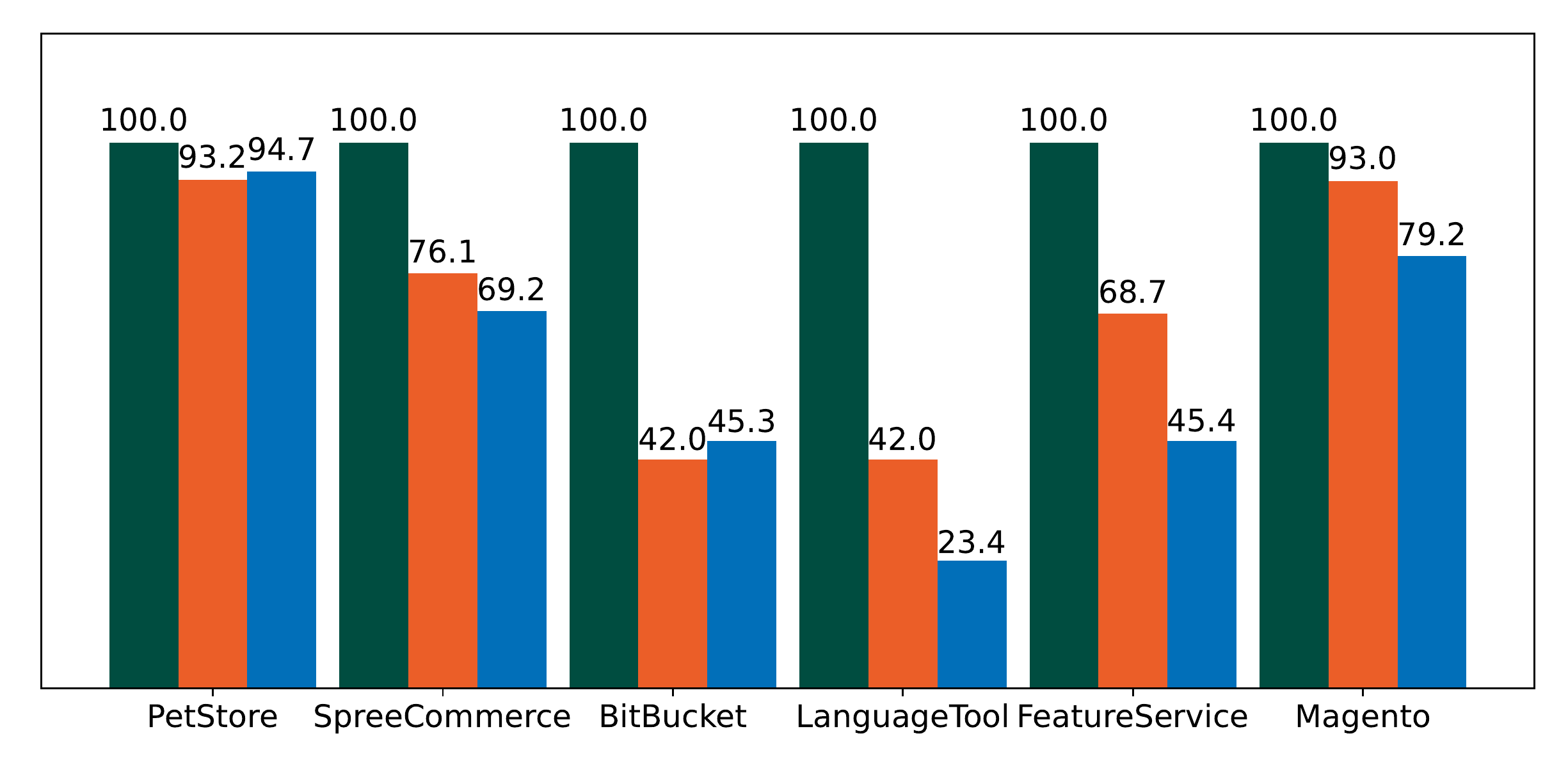}
    \caption{Normalized average code coverage}
      \label{fig:bar-code-coverage}

  \end{subfigure}
  
      \begin{subfigure}[b]{\linewidth}
    \centering
    \includegraphics[width=\textwidth]{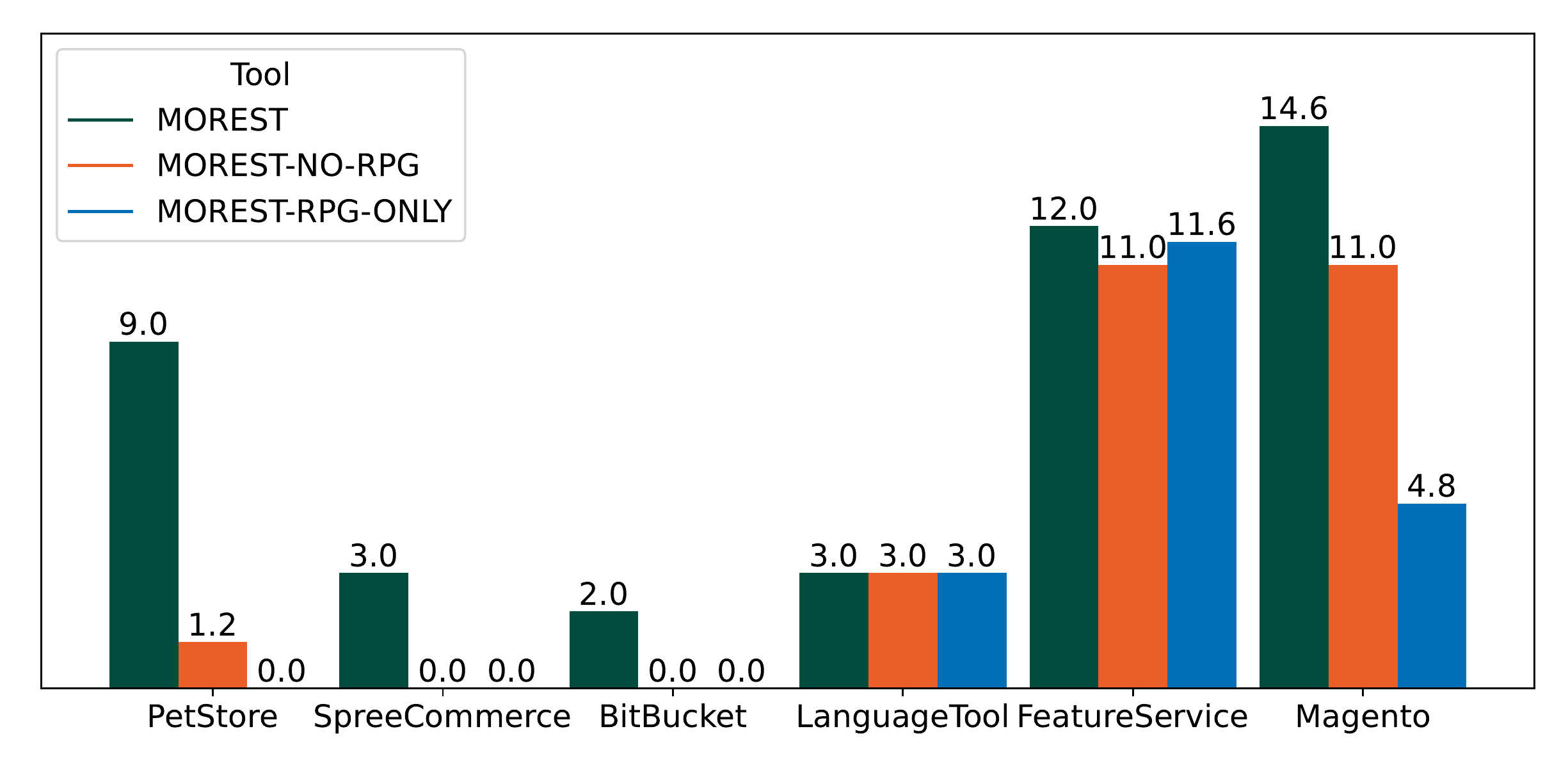}
    \caption{Average unique bug detection}
          \label{fig:bar-bug}

  \end{subfigure}
  
  \caption{The performance of \tool{}, \toolnoatm{} and \toolatmonly{} on both normalized average code coverage ($\mu LOC$) and bug detection.}
  \label{fig:bar}
\end{figure}
\subsection{Ablation Study (RQ3)}
To investigate how \atmab{} and dynamic \atmab{} updating contributes to boosting the testing effectiveness of \tool through high-level guidance and adaptive updating, we perform a ablation study on each component. To study the contributions separately, we implement two variants of \tool as following (1) \toolnoatm by disabling \atmab{} guidance, (2) \toolatmonly by removing the dynamic \atmab{} updating part. The results are averaged using five runs (with time budget one hour) to avoid the statistics bias. \figu~\ref{fig:bar-code-coverage} presents normalized code coverage of baselines on our datasets (To better visualization and understanding, we normalize the code coverage by $normalized\_coverage = \frac{LOC}{LOC(\tool)}$). \figu{}~\ref{fig:bar-bug} shows the number of detected bugs.

In general, as shown in \figu{}~\ref{fig:bar}, we observe that \tool outperforms \toolatmonly and \toolnoatm  on both code coverage and bug detection. In specific, we take Petstore as an example. To cover more operations related to the 'Pet' schema, an instance of pet should be firstly created by \textit{addPet}. However, \toolnoatm fails to generate such call sequences, which restricts further exploration during testing. Additionally, properties named \textit{status}, within both 'Order' (refers to the status of order) and 'Pet' (refers to the status of pet) schemas, leading to invalid producer-consumer dependencies in \atmab{}. The performance of \toolatmonly{} is limited by such cases. Thus, we can infer that the \atmab{} guidance and dynamic \atmab{} updating can both boost the effectiveness of the \rstapi testing process.

Besides, as we can see from \figu{}~\ref{fig:bar}, \tool achieves better performance other approaches regarding both code coverage and bug detection. In particular, \tool generates call sequences with the guidance of \atmab{} and dynamically updates \atmab{} to adaptively improve call sequences to obtain higher code coverage, which increasing the possibility to detect bugs.

\subsection{Threats To Validity}
The first internal threat comes from the choices of configurable options in the design.
Currently, we empirically set the values for these options and the actual values of the options used in the experiments can be found on our website~\cite{morest}.
Although the choice of configurable options can affect the performance of \tool{}, the experiment results can demonstrate that at least with the current set of option values, \tool{} can outperform state-of-the-art techniques.
Therefore, we leave the fine-tuning of these options as future work.
Another internal threat is that the current RPG model can only reflect the APIs documented in the OpenAPI specifications.
For undocumented APIs, a possible solution is to infer their related schemes in an online manner by analyzing the feedback of the testing process.
Another possible solution is to use client-side analysis techniques~\cite{js-dynamic,js-static,web-testing} to generate traffic between clients and servers and analyze the traffic to infer the correct usage of the undocumented APIs.
We leave the support of undocumented APIs as future work.

The external threats mainly come from the experiment settings.
The testing techniques evaluated in the experiments are random by nature.
To mitigate the random factors, we repeated each experiment for five times and conducted statistical tests.
Therefore, the experiment results are statistically sound.
To address the generality concern, we chose a diverse dataset consisting of six \rstservice{}s with various sizes and features.
Moreover, we chose open source \rstservice{}s so we can perform in-depth experiment result analyses via scrutinizing the code coverage and classifying the detected failures into unique bugs.
Last but not the least, to improve the fairness for technique comparison, we gave each technique a generous time budget of 8 hours while in \cite{restler} the longest time budget is 5-hour and in \cite{resttestgen} the time budget is 0.5-hour.




\section{Related Work}
Instead of discussing all related works, we focus on the \rstservice{} testing techniques, SOAP service testing techniques and model-based testing techniques.

\header{Blackbox \rstservice{} testing techniques.}
Several blackbox techniques were proposed to generate meaningful call sequences.
\restgen{}~\cite{resttestgen} builds Operation Dependency Graphs (ODGs) with the OpenAPI specifications to model \rstservice{}s and crafts call sequences via graph traversal.
The quality of the generated call sequences is limited by the quality of the OpenAPI specifications which directly affect ODG building.
Meanwhile, \restler{}~\cite{restler} builds call sequences with a bottom-up approach which starts with single operation call sequences and gradually extend the call sequences by appending more operations after trial and error.
Comparing to these techniques, \tool{} can enjoy both high-level guidance and the flexibility of dynamic adjustments to achieve better performance.

Other than call sequence generation, some blackbox techniques focus on 
improving the operation input parameter generation~\cite{grammar-fuzzing-1,grammar-fuzzing-2,restler2}.
These techniques utilize predefined inter-parameter constraints, input grammar, or mutators to generate diverse input parameters for the test cases.
Input parameter generation is orthogonal to call sequence generation, we plan to adopt advanced input parameter generation in the future.

\header{Whitebox \rstservice{} testing techniques.}
\evomaster{}~\cite{evomaster} is a whitebox \rstservice{} testing technique which instruments the target \rstservice{} and monitors its database to collect useful execution feedback and data to guide the evolutionary algorithm based test case generation.
Comparing to the the blackbox techniques, on the one hand, \evomaster{} is more effective in testing deeper logic inside the \rstservice{} since it can collect and use more information about the target service to guide the test case generation.
On the other hand, \evomaster{} can only instrument Java/Scala/Kotlin based \rstservice{}s and requires access to the database, limiting its application to closed source projects or projects implemented with other programming languages.


\header{SOAP Testing Techniques}
Simple Object Access Protocol (SOAP) is a standards-based web service access protocol first proposed by Microsoft.
Some previous works focus on testing SOAP~\cite{soap-root} based web services~\cite{penta-survey, penta-survey-2, harman-survey}.
Specifically, some of them use Web Services Description Language (WSDL) specifications to guide the testing process~\cite{wsdl-1, wsdl-2, wsdl-3, wsdl-4, wsdl-5, wsdl-6}.
Despite the similarities, REST is proposed to address the shortcomings of SOAP and is gaining more popularity in recent years~\cite{rest-root}.
The fundamental service  interaction models in REST and SOAP are different.
Therefore, testing \rstservice{}s requires different strategies.


\header{Model-based testing techniques.}
Besides the techniques for \rstapi{} and SOAP service testing, model-based testing techniques~\cite{FSE2019-WebTest,biagiola2017search,mesbah2011invariant,yu2015incremental,web-testing} are also related to \tool{}.
In general, these techniques use models, say a finite-state machine~\cite{test_fsm}, to describe the system-under-test and generate tests by covering different states in the model~\cite{model_testing_survey}.
In particular, Palulu~\cite{palulu} is a technique developed based on Randoop~\cite{randoop}, which uses a call-sequence model to quickly generate legal but behaviorally-diverse tests for Java classes.
Palulu's intuition of using a model to guide the generation of valid API call sequences is similar to \tool{}'s.
However, the difference is that the model used in \tool{} is tailored for \rstapi{} and it is dynamically updated during testing.

\section{Conclusion}
In this paper, we propose \tool{} --- a model-based blackbox \rstapi{} testing technique.
\tool{} learns from the OpenAPI specifications to build a \atm{} (\atmab{}), which encodes both schema and API operation information.
\tool{} can use \atmab{} to provide high level guidance for call sequence generation and continuously refine the \atmab{} with execution feedback.
We evaluated \tool{} on 6 open source \rstservice{}s and the results showed that \tool{} can significantly outperform state-of-the-art techniques in both coverage and bug detection. 

\section*{Acknowledgements}
This research is partially supported by the National Research Foundation, Prime Ministers Office, Singapore under its National Cybersecurity R\&D Program (Award No. NRF2018NCR-NCR005-0001), NRF Investigatorship NRFI06-2020-0022-0001,  the National Research Foundation through its National Satellite of Excellence in Trustworthy Software Systems (NSOE-TSS) project under the National Cybersecurity R\&D (NCR) Grant award no. NRF2018NCR-NSOE003-0001.
This research is supported by the Ministry of Education, Singapore under its Academic Research Fund Tier 3 (MOET32020-0004). Any opinions, findings and conclusions or recommendations expressed in this material are those of the author(s) and do not reflect the views of the Ministry of Education, Singapore.


\bibliographystyle{plain}
\bibliography{references}
\end{document}